\documentclass[10pt, journal, compsoc]{IEEEtran}
\ifCLASSINFOpdf
\else
\fi
\usepackage{hyperref}
\usepackage{booktabs}
\usepackage{amsmath}
\usepackage{amssymb}
\usepackage{graphicx}
\usepackage{multirow}
\usepackage{color}
\usepackage{wasysym}

\def\TN{{\texttt{CS}}}
\def\MI{{\texttt{SEC}}}
\def\PRO{{\texttt{PRO}}}
\def\MIPRO{{\texttt{SEC+PRO}}}

\def\CVSS{{\texttt{CVSS}}}

\def\AV{{\texttt{AV}}}
\def\AC{{\texttt{AC}}}
\def\UI{{\texttt{UI}}}
\def\PR{{\texttt{PR}}}
\def\C{{\texttt C}}
\def\I{{\texttt I}}
\def\A{{\texttt A}}

\newcommand*\rot{\rotatebox{90}}

\begin{document}
\title{The Effect of Security Education and Expertise on Security Assessments: the Case of Software Vulnerabilities}

% \author{\IEEEauthorblockN{Luca Allodi}
% \IEEEauthorblockA{Eindhoven University of Technology\\
% Eindhoven, The Netherlands\\
% Email: l.allodi@tue.nl}
% \and
% \IEEEauthorblockN{Marco Cremonini}
% \IEEEauthorblockA{University of Milan\\
% Milan, Italy\\
% Email: marco.cremonini@unimi.it}
% \and
% \IEEEauthorblockN{Fabio Massacci}
% \IEEEauthorblockA{University of Trento\\
% Trento, Italy\\
% Email: fabio.massacci@unitn.it}
% \and
% \IEEEauthorblockN{Woohyun Shim}
% \IEEEauthorblockA{Korea Institute of Public Administration\\
% Seoul, South Korea\\
% Email: whshim@kipa.re.kr}}

\author{Luca~Allodi,
        Marco~Cremonini,
        Fabio~Massacci,
        Woohyun~Shim% <-this % stops a space
\IEEEcompsocitemizethanks{\IEEEcompsocthanksitem L. Allodi is at the Eindhoven Univ.\ of Technology, NL. (\url{l.allodi@tue.nl}).
\IEEEcompsocthanksitem M. Cremonini is at the Univ.\ of Milan, IT. \url{marco.cremonini@unimi.it}. 
\IEEEcompsocthanksitem Fabio Massacci is at the Univ.\ of Trento,IT. \url{fabio.massacci@unitn.it}.
\IEEEcompsocthanksitem W. Shim is with the Korea Institute of Public Administration, KR.}% <-this % stops an unwanted space
\thanks{}}


%Telephone: (800) 555--1212, Fax: (888) 555--1212}


\maketitle

\begin{abstract}
In spite of the growing importance of software security and the industry demand for more cyber security expertise in the workforce, the effect of security education and experience on the ability to assess complex software security problems has only been recently investigated.
As proxy for the full range of software security skills, we considered the problem of assessing the severity of software vulnerabilities by means of a structured analysis methodology widely used in industry (i.e. the Common Vulnerability Scoring System (\CVSS) v3), and designed a study to compare how accurately individuals with background in information technology but different professional experience and education in cyber security are able to assess the severity of software vulnerabilities. 
Our results provide some structural insights into the complex relationship between education or experience of assessors and the quality of their assessments.
In particular we find that individual characteristics matter more than professional experience or formal education; apparently it is the \emph{combination} of skills that one owns (including the actual knowledge of the system under study), rather than the specialization or the years of experience, to influence more the assessment quality. Similarly, we find that the overall advantage given by professional expertise significantly depends on the composition of the individual security skills as well as on the available information.
%Together these results suggest that developing a well-trained workforce in information security depends on a multitude of factors and on the ability to combine different skills, needed by the complex nature of many information security problems.

% Cyber security is an hot topic and several universities are offering 
% bachelor and master degrees in the field. 
% It is unclear whether such extensive security knowledge can make a substantial
% difference in the ability to correctly assess software vulnerabilities over normal CS 
% education. In this empirical study we ask MSc students of two different universities 
% one with a BSc in security and one with a BS in plain CS to assess several vulnerabilities 
% and compare their assessments against the expert scores of 
% the Common Vulnerability Scoring System (\CVSS) v3. 
% We find that security knowledge is important over only few dimensions of the
% assessment. Further, for a given estimation of the individual components
% the estimation of the overall score of the  \CVSS\ formula is not affected by 
%  the assessor's security background. 
\end{abstract}


% \fi
\IEEEpeerreviewmaketitle

\section{Introduction}
Given the raising importance of cyber security, several Universities have recently introduced security courses and degree programs~\cite{schneider2013cybersecurity,yue2016teaching}. Some governments already identified key knowledge areas for their workforce (e.g.~\cite{Curriculum:CAE}) whilst professional organizations also 
proposed curricula for Security IT management~\cite{Curriculum:ISACA} and cyber security at large~\cite{Curriculum:ACM}. 
However, the set of core skills is widely debated with opinions often split between teaching adversarial thinking or principles and abstractions~\cite{mcgettrick2013toward,conklin2014re,santos2017challenges}. Widely debated is also the role of governments through public initiatives mandating academic training as a formal requirement for recognized professionals~\cite{burley2014would,reece2015professionalisation}.

In this scenario, it is surprising that only few studies considered measuring the effectiveness of security education and professional experience with respect to relevant security tasks. 
%Instead, many studies are focused on the effectiveness of automatic tools, for instance to perform code analysis, vulnerability assessment, or intrusion detection. However, almost every seasoned security expert would admit that for successfully dealing with many security tasks, good tools must always be complemented with skilled personnel. Therefore, the lack of analysis about how personal skills contribute to address security tasks looks even more puzzling.
%
Experiments to quantitatively evaluate the benefits of education and experience with respect to the ability to tackle a cybersecurity problem have been mostly performed by considering specific technical skills, often the ability to write a secure code or to recognize programming errors. In one of those studies~\cite{edmundson2013empirical}, where the problem of finding vulnerabilities in code was considered, no correlation between education and code analysis effectiveness was found whilst a negative correlation emerged between the number of vulnerabilities found and analysts' years of experience in security. An hypothesis to explain these results is that complex relationships between professional knowledge and problem solving are at work~\cite{hardiman1989relation,dufresne1992constraining}.

%In this work, rather than focusing on specialized technical skills, we are interested in considering a multifaceted  security problem requiring multiple skills to be addressed. Such a situation is common in security operations and management when the person in charge is not a specialist of the specific security problem and, by nature, the problem would require a certain familiarity with different technical aspects, the evaluation of uncertain effects, and an attention to management aspects. 

\textbf{Goal and methodology.} Our long term objective is to understand how
different security education levels, professional experience, and personal skills 
affect the outcome of software security activities. The latter involve a combination of 
technical skills, system awareness, and management-level perspectives, forming 
the set of attitudes (or ``mindset'')~\cite{HOLM-2015-CS} that is often part of the job 
of information security professionals~\cite{Wash-2010-SOUPS} or 
managers~\cite{Chen-2011-MISQ}. This is also reflected in best practices for 
security education at large (e.g.\ the DHS table of minimal 
content~\cite{Curriculum:CAE}). 
%Whereas finding a complex task covering these aspects is not an issue (e.g. SOC 
%operation, privacy evaluations, ..), it remains difficult to evaluate its outcome over
% an objective basis for comparison, hence making it impossible to isolate the 
% effects of subject variables on the overall metric of reference. Reflecting this
%difficulty, most studies in the literature evaluate `simple' security tasks that cover 
%only one or few aspects of the general security problem (e.g. bug identification).
%
%In more details, assessing the severity of software vulnerabilities is a precondition for deciding risk treatment activities and to prioritize the allocation of resources for mitigating possible adverse effects. The severity of a vulnerability is typically the result of an analysis encompassing different perspectives, such as technical characteristics of the impact, required interactions with users, or attack vectors. 
%
Part of the challenge is to identify a task that is i) simple and structured enough to be amenable to controlled experiments whilst being ii) rich enough to provide a non obvious challenge to human participants to the experiments,  and iii) decomposable so that one could probe different parts of the security skill sets. 

We propose to use software vulnerability assessment
as a prima facie proxy to evaluate the interplay between different aspects of a 
security task: understanding the impact of a software vulnerability on a system 
requires technical and operational skills (e.g. to evaluate impact vectors), as well as 
user-oriented and a management perspective (e.g. to estimate user requirements 
on the attack, or consequences of an attack). The choice of vulnerability 
assessment has another significant advantage: it comes with a pre-defined, world-
standard evaluation framework that, for each of these aspects, provides an 
objective reference frame: the \emph{Common Vulnerability Scoring System} 
(\CVSS)~\cite{Mell-2007-CMU}~\footnote{\CVSS\ is the most employed software 
vulnerability evaluation methodology in industry and government 
\cite{Scarfone-2010-SCAP}. It is the result of the joint work of national standard 
bodies, software manufacturers, non-software companies, and Computer 
Emergency Response Teams (CERTs).}. It is of particular importance that its usage 
in corporate settings is not targeted to software security specialists. It is a tool that 
`general security practitioners' with a CS background should be able to operate. 
For instance, the User Interaction (\UI) metric only requires the assessor to 
comprehend how the vulnerable software is used and in particular whether or not a 
user's action is required for the exploit to materialize; the Attack Vector (\AV) metric, 
instead, requires evaluating aspects of the network stack of the vulnerable 
application.  Obviously, such task is only a part of the skills of a software security 
experts. Security testing is another example of activities and related skills. The DHS 
Curriculum provides a good starting point for a list of such activities and skills set 
\cite{Curriculum:CAE}. However, as one cannot test all them at once (too many 
confounding factors) we decided to focus on the assessment task for this study.

To investigate the relation between skills and quality of a software security task 
represented by a vulnerability assessment, we conducted an experiment with three 
groups of participants, asking them to score a set vulnerabilities following \CVSS\ 
guidance. Our participants were both students and professionals, the former divided 
between those with or without specific security education, the latter with several 
years of experience in the security field but no security education at academic level.   

%The three groups of people involved in the experiment (total $n=73$ participants) have been selected having different levels of security training and years of experience (35 students in a computer science major, 19 students in an information security major, and 19 security professionals with several years of experience but no academic security education). Some participants knew what \CVSS\ is used for and its scores associated to CVE vulnerabilities, but none had experience with \CVSS \texttt{v3} vulnerability assessment or knew the specific metrics used to produce the score. 
%
%The experiment has been organized by replicating the procedure performed by the \CVSS\ Special Interest Group (SIG) for its own operations by asking each participant to complete 30 vulnerability assessments in 90 minutes by using only the CVE vulnerability description as the only technical information available. The official assessments produced by the \CVSS\ SIG were used as our benchmark to compare between-group performance.

%Employing students in experiments with the goal of studying problem solving performance in professional contexts has a long tradition in cognitive and managerial studies and is mostly motivated by the difficulty of recruiting professionals. It has, however, often provided useful insights and even results almost equivalent to those produced by professionals, under certain circumstances ~\cite{host2000using,Salman-ICSE-15,kalyuga2003expertise}. 

\textbf{Main contributions.} 
%Unsurprisingly, we find that security knowledge is clearly beneficial to assessing the severity of a vulnerability (i.e., participants with security background score clearly better than those without it). % and, globally, participants scored clearly worse than SIG. 
A preliminary finding is that being competent in security, through education or experience, improves the ability to evaluate vulnerability severity. Albeit expected, the quantification of such improvement may represent the basis for cost/benefit analyses (Eg whether pursue a specialized but costly Master degree or a professional training).

Indeed, we found that experienced security professionals have no clear advantage over younger students with a security specialization. Other experiments in different fields have shown that expert performance is close to that of novices when solving problems in novel situations for which `precompiled' rules are not helpful~\cite{singh2002physical}. In an other case, {\em expertise reversal effect} has been detected when experienced learners decide to ignore novel but detailed instructional procedures in favor of `I-know-better' mental schemes~\cite{kalyuga2003expertise}. Hence, formal security education could produce a workforce able to compete with seasoned professionals when asked to perform well-formalized but relatively new security tasks.

%Further analysis of our results has suggested that a broad and diversified
% professional skill set is an advantage to tackle certain complex security problems,
% and that a poor or ambiguous definition of the security problem could severely 
% affect the assessment accuracy.

%Clearly, considering the limits of our study and the lack of extensive research on this issue, our findings can be used just to reinforce hypothesis to be further tested, rather than be considered as proofs. However, 
Whilst this result is encouraging, we believe that more studies and tests are needed in order to establish evaluation methodologies and metrics for advanced education and professional training that often are considered as important investments.

\section{Related Work} \label{sec:related:work}

\subsection{Information Security Education}
Education is a recurrent topic in information security, often related to the definition of guidelines for academic curricula needed by a well-trained workforce or considered in field studies comparing the performance of different groups of professionals when faced to a security problem.

Relevant initiatives for the definition of curricula and guidelines have been carried out by professional associations like ISACA~\cite{Curriculum:ISACA} or ACM~\cite{mcgettrick2013toward} as soon as cybersecurity has been recognized as an emergent IT profession. The most recent initiatives exhibit a remarkably improved maturity of proposals. In particular, a joint initiative between ACM, IEEE, AIS, and IFIP has produced a detailed framework for cybersecurity curriculum encompassing several knowledge areas~\cite{Curriculum:ACM}. The NSA, instead, together with the Department of Homeland Security, has been supported for years some Centers of Academic Excellence in Cyber Defense (CAE-CD) and, in connection with that initiative, the definition of a list of cybersecurity Knowledge Units (KUs) to be mapped on specific topics included into academic curricula~\cite{Curriculum:CAE}.    
%
%operational mistakes that end-users or developers do during their day-to-day activity or with respect to the lack of a well-trained workforce as needed by industry.
%Related to our study are works of the second type, which discuss the need of specific security education and analyze the characteristics that such training should have with respect to the goals.
%For example,~\cite{mcgettrick2013toward} summarizes the outcome of a workshop organized by the ACM's Education
%Board and sponsored by the US National Science Foundation (NSF) aimed at providing guidelines for new academic courses focused on information security. 
%Overall, they find that ideally an academic program should provide a strong Computer Science background and in the remaining time should address a variety of security-related topics, both theoretical and practical. Therefore, different curricula could be envisaged and some general principle should be enforced, like multidisciplinarity, a security mindset and the need to mix theory with hands-on activity.
%
With respect to the content of a cybersecurity curriculum, both the Joint Task Force and the CAE-CD propose a broad spectrum of competences, including technical and non-technical KUs (i.e., CAE-CD), or human, organisational, and societal security knowledge areas in addition to the more traditional system, data, software, component, and connection areas (i.e. the Joint Task Force). Other analyses have further contributed to the discussion~\cite{schneider2013cybersecurity,Manson:2014,yue2016teaching,santos2017challenges}.

%The difficulty of measuring the effectiveness of security education with respect to clearly defined and verifiable industry's quality levels for information security is also a central problem~\cite{conklin2014re}.
%,
%More oriented to professionals rather than to education, some works have discussed the trend towards professionalization of information security, meaning a single occupation category with certifications, skill-based exams, licensures, and typically requiring academic graduation for entry
Relative to the professionalization of cybersecurity, some have discussed the difficulty that professionals have to agree on a set of well-established characteristics and the need for the cybersecurity profession to evolve~\cite{burley2014would,reece2015professionalisation,conklin2014re}. 
These analyses are connected to our work because they reflect the lack of experimental studies on the effectiveness of security education and security experience, and on the emerging scenario composed by many security profiles and professions. 
%Knowles et al.~\cite{knowles2017all} looks at the criteria used in practice of professional cybersecurity qualifications, and found discrepancies between the criteria perceived usefulness and their adoption. 

Several works obtained results from controlled experiments with students and professionals performing security tasks, although almost all of them focused on specific skills or even specific tools.
In~\cite{wermke2017security}, a sample of developers recruited from GitHub was asked to complete simple security programming tasks. It turns out that the only statistically significant difference was determined by the years of programming experience, which indicates that familiarity with the task is the main driver, rather
than professional or educational level. This is in line with results in~\cite{edmundson2013empirical}, whereby security professionals do not outperform students in identifying security vulnerabilities. The usability of cryptographic libraries has been studied in~\cite{acar2017comparing}. Relevant for our work is the fact that different groups of individuals, with different levels of education or experience, have been recruited. They found that the participant’s programming skill, security background, and experience with the given library did not significantly predict the code's overall functionality. Instead, participants with security skills produced more secure software, although neither as good as expected nor as self-evaluated by participants. We have found compatible results, although under very different settings and more nuanced. All these works differ from ours in studying the performance of individuals with respect to a specific technical security skill or tool rather than studying how education, experience, and the combination of skills correlate with more general security problem solving performances. The one closer to us is~\cite{acar2016you}, where Android developers' decision making performance are also analyzed with respect to education and experience. The experiment was based on observing how developers with different background perform when provided with different types of documentation and it found a sensible difference between professionals and students. However, with respect to our work, the analysis based on the different backgrounds was limited and only a minor part of the study. 

In other research fields, namely human-computer interaction and software engineering, there has been an ample discussion on the accuracy and precision of problem solving by different types of evaluation panels. Some studies have since long investigated the factors that distinguish experts from novices~\cite{sweller1988cognitive} or in evaluating software usability~\cite{HERTZUM-2001-IJHCI,LING-2009-IJHCI}. In general, what studies in different fields have demonstrated is that, under specific circumstances, experts not necessarily perform better than novices in technical tasks, and that the relation between knowledge and problem solving ability is often unclear. These observations further motivate a specific investigation for security problems.

\subsection{Vulnerability assessment}
Vulnerability assessment represents a fundamental phase for security risk analysis and the prioritization of activities. 

In this regards, it is both a technical task and a security management issue, which includes the adoption of a structured analysis methodology supporting a metric for assigning a qualitative or quantitative evaluation to vulnerabilities and the availability of assessors with adequate skills and training~\cite{HOLM-2015-CS,Scarfone-2010-SCAP,calder2008governance}. Vulnerability distributions have been studied 

Related to our work, some research analyzed the distribution of vulnerability assessments performed by means of the \CVSS~\cite{SCAR-MELL-09-ESEM,LIEU-ZHANG-10-CC,Gallon-2011-NTMS}. Similarly, \cite{allodi-2012-BADGERS, Allodi-13-IWCC,Allodi-2014-TISSEC} investigated whether vulnerabilities with a high \CVSS\ score corresponds to an equally high risk of exploitation in the wild.

Some studies have considered the difference in vulnerability scoring produced by assessors with different profiles~\cite{Hibshi-2015-RE,HOLM-2015-CS}. 
%Confirming previous results in related research \cite{BOZORGI-etal-10-SIGKDD,Christey-2013-BHUSA,zimmermann2010makes}, 
 %they find that additional information, with respect to that provided by the \CVSS\ base metric, is required to estimate vulnerability risk.  
%The additional information identified by experts in \cite{HOLM-2015-CS} is largely included in extensions of the \CVSS\ base metric, namely the Temporal and Environmental metrics \cite{Mell-2007-CMU,first-2015-CVSS3}.
Differently from these works, we aim at studying to what extent security education and practical experience influence the outcome of a vulnerability assessment. %In order to perform such an experiment, we adopted the recent \CVSS\ v3 as our reference methodology because of its clear structure and applicability to a variety of security scenarios. %Further, ours is the first study that tests the use of \CVSS\ v3 in scoring software vulnerabilities in the software and security engineering literature.

%\section{Hypotheses} \label{sec:hypothesis}
\section{Study Design} \label{sec:study}

%We designed a study to compare how effectively students that received a formal education in information technology and professionals with experience in cybersecurity perform a vulnerability assessment, according to a well-established methodology: \CVSS\ v3. 

\subsection{Analysis goals and research questions}
%The analysis is aimed at evaluating the effect of formal education and of professional experience on the ability to perform a security assessment. More generally, we intend to investigate how a sample of students with a formal training in cybersecurity and a sample of professionals with experience in cybersecurity perform in a task like vulnerability assessment according to \CVSS\ methodology, which requires an heterogeneous set of skills and presents uncertainties (i.e., a situation very common in cybersecurity).
%To illustrate this, we define two sub-goals:
In this study we evaluate the effect of different subject characteristics on \emph{technical, user and system-oriented, \emph{and} managerial} aspects of a security task. Specifically, our study aims at the following two goals:

\subsubsection*{Goal 1: Effect of security knowledge} We first evaluate
the effect of skills in security on the accuracy
of an assessment. We distinguish between skills acquired through \emph{formal security education}, meaning academic-level specialized security courses, and through \emph{professional experience}, as years of work in the security field. 
%Typically, the expectation is that skills brings better evaluations, but, as already anticipated, that seems not always the case~\cite{edmundson2013empirical,singh2002physical}: The nature of the security problem and the familiarity of assessors may change the performance. 
%One typical expectation is that accuracy of an assessment generally increases with security expertise; yet, it is unclear whether security expertise is a significant factor for \emph{all} aspects of a security assessment, or if it influences only a few (and by how much).
%Furthermore, it remains unclear to which extent the professional expertise improves a security assessment over academic education. The aim here is to identify how security experts `by training' perform with respect to the quality of an assessment, and whether their performances are comparable to those of security experts `by profession'. This will guide the discussion on the importance of security education and its limitations.
We formulate two research questions:
\begin{quote}
\textbf{RQ1.1}: Does formal security education have a positive impact on the accuracy of a structured security assessment?\\
\textbf{RQ1.2}: Does professional expertise systematically improve assessment accuracy over formal security education?
\end{quote}

\subsubsection*{Goal 2: Effect of specific security competences} The second aspect we want to investigate is how specific skills impact the accuracy of the security assessment. Our underlying hypothesis here is that the analysis should consider the specific mix of competences, derived from education and professional experience, in order to identify meaningful relations between education/experience and problem solving ability in security. The provision of a `standardized' portfolio of security competences has been pushed by both academic curricula~\cite{mcgettrick2013toward} and industry~\cite{von2005information}, yet it is unclear whether this fits
well the requirements of a real world scenario. 
%For example, specific knowledge on system and data security may affect assessment accuracy on aspects of attack impact and consequences, whereas previous experience with software attacks may help in correctly identifying measure of attack likelihood or complexity. The aim of this research goal is to evaluate the effect of combination of skills on the accuracy of a security assessment to provide indications on the specific skill-sets required and their effect on specific aspects of an assessment.
This research goal addresses the following research questions:
\begin{quote}
\textbf{RQ2.1}: Does the mix of competences affect the overall accuracy of an assessment?\\
\textbf{RQ2.2}: Which specific aspects of a security assessment are more influenced by which specific skill?
\end{quote}

\subsection{Task mapping and vulnerability selection}

The \CVSS\ v3 framework 
provides a natural mapping of different vulnerability metrics on aspects of the larger spectrum of security competencies we are considering: technical, user-oriented, and management-oriented. Using \CVSS, the assessor performs an evaluation of the vulnerability based on available information. %: number of dimensions over which a vulnerability is assessed based on 
Table \ref{tab:CVSSbasemetrics} provides a summary of \CVSS's Base metrics used in this study, their possible values, and their relation with the three competency levels we identify.
\begin{table*}[t]
\centering
\small
\caption{Summary of considered \CVSS\ v3 Base metrics and mapping to competency levels.}
\label{tab:CVSSbasemetrics}
\begin{tabular}{ l p{0.15\textwidth} p{0.25\textwidth} p{0.1\textwidth} p{0.35\textwidth}}\toprule
% \multicolumn{4}{c}{\textbf{Exploitability metrics}} \\
 \CVSS\ & Metric & Metric desc. & Values & Skill set\\ \midrule
\AV & Attack Vector & Reflects how remote the attacker can be to deliver the attack against the vulnerable component. The more remote, the higher the score. & \texttt{Physical}, 
\texttt{Local}, \texttt{Adjacent Net.}, \texttt{Network}. &\multirow{2}{0.3\textwidth}{The assessor understands the technical causes and vectors of attack related to a software vulnerability. This encompasses knowledge of vulnerable configurations, local and remote attack delivery, and aspects related to attack engineering.}  \\
\AC & Attack Complexity & Reflects the existence of conditions that are beyond
the attacker's control for the attack to be successful. & \texttt{High, Low.} &\\
\midrule
\PR& Privileges Required & Reflects the privileges the attacker need have
on the vulnerable system to exploit the vulnerable component. & \texttt{High, Low, None.} & \multirow{2}{0.3\textwidth}{The assessor understands the interaction between the vulnerable system, the user, and the attack. For example, attacks against administrative users may require specific attack techniques (e.g. spear-phishing); similarly, user behaviour may affect the outcome of a security problem (e.g. ignoring alert dialogues).} \\
\UI&User Interaction & Reflects the need for user interaction to deliver a successful attack. & \texttt{Required, None.} \\ [0.57in]
% \multicolumn{3}{c}{\textbf{Authorisation Scope metrics}} \\\hline
% Metric & Description & Values\\\hline
% Scope & Defines whether the impacted component is under the same authorisation scope as the vulnerable component. & Unhanged, Changed. \\[0.1ex]
% \multicolumn{4}{c}{\textbf{Impact metrics}} \\
% ID & Metric & Description & Values\\\hline
\midrule
\C & Conf. & Measures the impact to the confidentiality of information on the impacted system. & \texttt{None, Low, High.} &\multirow{3}{0.3\textwidth}{The assessors can evaluate the repercussions of a security problem over business-level aspects such as data exfiltration and system performance.} \\
\I&Integrity & Measures the impact to the integrity of information stored on the impacted system. & \texttt{None, Low, High.}\\
\A&Availability & Measures the impact to the availability of the impacted component. & \texttt{None, Low, High.}\\\bottomrule
\end{tabular}
\end{table*}

%\subsection{Vulnerability selection}
To guarantee the vulnerabilities' representativeness the wider set of vulnerability characteristics, we chose the vulnerabilities for our experiment by randomly sampling thirty vulnerabilities from the one hundred used by the \CVSS\ {\em Special Interest Group} (SIG) to define the \CVSS\ standard. This also assures that the sample is representative of the distribution of \CVSS\ vulnerability measures in the population of vulnerabilities (which are not uniformly distributed, see for example~\cite{SCAR-MELL-09-ESEM,Allodi-2014-TISSEC}). The vulnerability descriptions are taken from the National Vulnerability Database (NVD), the reference dataset of disclosed software vulnerabilities. 

% {\color{red} add part on vuln description, cwes, and aggregate of cvss metrics values to show representativeness}

\subsection{Participants and recruiting procedure}

We follow~\cite{MEYER-1995-JBES} and performed a natural experiment recruiting three groups of individuals (total $n=73$ participants): 35 major students with no training in security; 19 major students with three to four years of specific training in security; 19 security professionals with several years of experience. Some participants knew what \CVSS\ is used for and its scores associated to CVE vulnerabilities, but none had experience with \CVSS \texttt{v3} vulnerability assessment or knew the specific metrics used to produce the score. 

The experiment has been organized by replicating the procedure performed by the \CVSS\ Special Interest Group (SIG) for its own operations by asking each participant to complete 30 vulnerability assessments in 90 minutes by using only the CVE vulnerability description as the only technical information available. The official assessments produced by the \CVSS\ SIG were used as our benchmark to compare between-group performance.

% Students of a Computer Science MSc Major with no training in security aspects of IT; students of an Information Security MSc Major with three to four years of specific training in security; security professionals, among which chief information officers (CISOs), security consultants, security auditors, and CERT team leaders. 
Unfortunately, recruiting subjects with very different profiles makes it hard to control for possible confoundings; for example, some professionals may have received an education equivalent to that of (a group of) student subjects, or some students may have changed masters during their student career. As these effects are impossible to reliably measure, we explicitly account for the (unmeasured) in-subject variability in the analysis methodology and report the corresponding estimates.

\subsubsection{Students} Students participating in our study are MSc students of two European Universities, both requiring proficiency in English and a background in computer science. The first group, \MI, is enrolled in the Information Security MSc of the University of Milan%\footnote{University names are omitted for compliance to submission anonymity.} 
that completes a BSc in Information Security held at the same university. %~\footnote{The University of Milan has a BSc and a MSc in Information Security since 2003/2004.}. 
% \MI\ students know the fundamentals of software vulnerabilities and exploits, as well as of risk analysis. Students in this group have a 3 years BSc degree on computer security, which includes access control, cryptography, network and system security, and privacy\footnote{The education curriculum amounts to 120 ECTS where 1 ECTS is 25 hours of work for the student in variable proportion between 
% individual study, lectures and supervised.}.
The second group, \TN\ group, is composed of students enrolled in a Computer Science MSc at the University of Trento, Italy. 
\MI\ subjects were recruited during the \emph{Risk Analysis and Management} course at the first year of their MSc; \TN\ students were recruited during the initial classes of the course \emph{Security and Risk Management}, the first security-specific course available in their MSc curriculum. %\TN\ students joined  a 3 years BSc degree on computer science.

Table~\ref{tab:CASknowledge} 
\begin{table*}[t]
\centering
\small
\caption{Core Knowledge Units for \TN\ and \MI\ students}
\label{tab:CASknowledge}
\begin{tabular}{l c c c c c c c c c c c c c c c c c}
%\hline
% \multicolumn{4}{c}{\textbf{Exploitability metrics}} \\
  & \rot{Basic Data Analysis} & \rot{Basic Scripting} & \rot{IT Sys. Components} & \rot{Networking Concepts} & \rot{Sys Admin.} & \rot{DataB. Mngmt. Sys.} & \rot{Net. Techn. and Prot.} & \rot{Op. Systems Concepts} & \rot{Prob. and Stats.} & \rot{Programming} & \rot{Fund. of Sec. Design} & \rot{Fund. of Crypto.} & \rot{Cyber Defense} & \rot{Cyber Threats} &  \rot{Net Defense} & \rot{Policy/Ethics Compl.} & \rot{Fund. of Inf. Assur.} \\\hline
\TN & \CIRCLE & \CIRCLE & \CIRCLE & \CIRCLE & \CIRCLE & \CIRCLE & \CIRCLE & \CIRCLE & \CIRCLE & \CIRCLE & \Circle & \Circle & \Circle & \Circle & \Circle & \Circle & \Circle \\
\MI & \CIRCLE & \CIRCLE & \CIRCLE & \CIRCLE & \CIRCLE & \CIRCLE & \CIRCLE & \CIRCLE & \CIRCLE & \CIRCLE & \CIRCLE  & \CIRCLE & \CIRCLE & \CIRCLE & \CIRCLE & \LEFTcircle & \Circle\\
\hline
\end{tabular}
\end{table*}
reports students skills as core knowledge units (taken from the U.S. Center for Academic Excellence (CAE) Core Knowledge 
Units in Cyber-security) of respective BSc programs. In particular, we see that they share some core Computer Science competences, whereas only the \MI\ group has been trained on core Information Security competences.

%  the  that both groups already completed at \MI's and \TN's respective universities to the U.S. Center for Academic Excellence (CAE) Core Knowledge Units in Cyber-security, a reference framework for information security education. \MI\ courses fully cover the majority of core topics. On the contrary, \TN's courses have a very limited coverage and only partially touch on critical aspects aspects of IT security.
%Security skills in the first group (\MI) can therefore be expected  to cover security fundamentals and some specializations, principles of risk analysis, and overall are oriented towards a `security mindset', that is a familiarity with regard to security problems and technologies. \TN\ students can, conversely, only be considered familiar with information technology and with a lively technological environment fostering innovation. 

%The two universities are consistently highly ranked in the fields of Information Engineering, so we assume that the relative quality of education at the two institutions is not a factor in the student's performances. The \MI\ group comprises 19 subjects, and the \TN\ group 35. Two students in \TN\ did not perform valid assessments for any of the tasks, dropping the number of valid subjects to 33 for this group.

\subsubsection{Professionals} Subjects in the \PRO\ group are members of a professional security community lead by representatives of the Italian headquarters of a major US corporation in the IT sector. 
Participants in our study have been recruited through the advertisement in the Community's programme of a training course on \CVSS\ v3. Participants in the \PRO\ group have different seniority in security and all professional profiles focus on security-oriented problems, technologies, and regulations. 
%The \PRO\ group comprises 19 professionals. 

To characterize \PRO\ experiences, we asked them to complete a questionnaire detailing
job classification and years of experience, education level, experience in vulnerability assessment, and expertise level in system security/hardening, network security, cryptography, and attack techniques. 
%It is useful to notice that these aspects reflect the domains of competences in information security identified by the CAE in Table~\ref{tab:CASknowledge}.
Of the 19 components of the \PRO\ group, 13 completed the questionnaire. The median subject in the \PRO\ group has six years of expertise in the security field, and roles comprise Security Analysts, CERT members, Pentesters and IT auditors. A detailed characterization of \PRO\ subjects over the other dimensions is given in Sec.~\ref{sec:subjvuln}.

%None of the participants in the three groups has received a formal training or attended a specific presentation of \CVSS\ v3 before the experiment.
%Overall, the study includes $n=71$ subjects, 33 in the \TN\ group, 19 in the \MI\ group, and 19 in the \PRO\ group. 
%All subjects in and \MI\ and \PRO\ finished the full assessment, whereas seven subjects in \TN\ did not. %This may be a preliminary indication that \TN\ students may have found on average the scoring exercise harder.

% The benefit of comparing the three groups based on emph{natural experiment} is that a group distinction based on the students' security knowledge obtained from 3 years of laboratory activities would be much more informative than 1 to 2 days of security training based on randomized trial.  

%Although students in both groups have different security knowledge and backgrounds, no group had previously performed vulnerability assessments using \CVSS\ (v1 or v2). 

\subsection{Data collection}

%\subsubsection{Experiment description} 

%\subsubsection{Experiment execution} 

% Subjects were asked to evaluate a set of vulnerabilities following \CVSS\ v3 guidelines after attending an identical introductory seminar explaining the metrics outlined in Section~\ref{sec:cvss}. 

Ahead of the experiment, participants attended an introductory seminar to \CVSS\ v3 held by one of the authors with several years of expertise on \CVSS. Content and delivery of the seminar have been identical for the three groups. After that, participants were given a printed sheet in tabular form reporting vulnerability description, \CVSS\ metrics, and requesting an evaluation on the confidence in the assessment. All participants had at hand a summary description of \CVSS\ v3 metrics obtained from the First.org's website for reference during the exercise. Time for completing the test was 90 minutes, chosen on the basis of previous pilot studies. All participants completed the assessment in the assigned time with the exception of seven students in the \TN\ group. All experiment material is provided for consultation at~\url{https://github.com/cvssexp/cvssmaterial}.

Table~\ref{tab:assessmentexample}
\begin{table}
\centering
\small
\caption{Example of assessment by one randomly selected participant for each \TN, \MI, \PRO\ groups compared to SIG's evaluation for CVE 2010-3974}
\label{tab:assessmentexample}
\begin{minipage}{0.95\columnwidth}
\footnotesize
Excerpt of the CVE 2010-3974: \emph{fxscover.exe in the Fax Cover Page Editor in Microsoft Windows XP SP2 and SP3, Windows Server 2003 SP2, Windows Vista SP1 and SP2, Windows Server 2008 Gold, SP2, R2, and R2 SP1, and Windows 7 Gold and SP1 does not properly parse FAX cover pages, which allows remote attackers to execute arbitrary code via a crafted .cov file, aka ``Fax Cover Page Editor Memory Corruption Vulnerability''.}
\end{minipage}

\begin{tabular}{p{0.15\columnwidth} lllllllp{0.15\columnwidth}}
\toprule
&\multicolumn{6}{c}{\CVSS\ assessment}&\\
\cmidrule{2-8}
Participant in group:&\AV&\AC&\PR&\UI&\C&\I&\A&Confident\\
\midrule
\TN&N&H&L&N&L&L&L&Yes\\
\MI&L&L&L&R&H&H&H&Unsure\\
\PRO&L&L&N&N&H&H&H&Yes\\
\midrule
\texttt{SIG}&L&L&N&R&H&H&H&- \\
\bottomrule
\end{tabular}
\end{table}
reports an example of vulnerability assessment. The answers from one participant, randomly chosen, for each group, are shown together with reference evaluations produced by SIG (bottom row). In this particular case, the \TN\ student had all answers wrong and, despite this, declared to be confident in his/her evaluation. Both the \MI\ student and the \PRO\ professional, instead, made one mistake, but exhibited different degree of confidence in their evaluation.

%The column {\em Confident} reports the self-reported confidence of that particular participant in scoring the vulnerability (as \emph{Yes, No, Unsure}). The two participants in the \TN\ and \PRO\ groups indicate good confidence, whereas the \MI\ participant is uncertain about the correctness of his/her assessment. The \TN\ participant, however, was wrong in evaluating the \C\I\A\ impact, assigning a \texttt{Low} score for each metric, and misunderstood both the attack vector (\AV) and the attack complexity (\AC) of the vulnerability. The \MI\ participant is the only one that correctly identified the requirement on \UI. Overall one can observe that for this vulnerability the \TN\ participant performed poorly, whereas the two \MI\ and \PRO\ participants performed almost identically (and correctly). This exercise was repeated for all thirty vulnerabilities.

%Subjects in the \PRO\ group were asked to answer the questionnaire ahead of the exercise.

%\subsubsection{Subject characteristics} We asked subjects in the \PRO\ group
%to answer a set of questions to detail their security profile. We asked six questions to define: job title; years spent in the current position; education level; 

\subsection{Analysis methodology}\label{sec:methodology}

We formalize a \CVSS\ assessment by assuming there exists a function $a_{i}(v_j)$ representing the assessment produced by assessor $i \in \{\TN \cup \MI \cup \PRO\}$ of vulnerability $v$ represented as the vector of \CVSS\ metrics to be evaluated ($j \in$ \{\AV, \AC, \UI, \PR, \C, \I, \A\}). 
We further define a function $e(a_{i}(v_j))$ that detects the error on metric $j$ by assessor $i$ on vulnerability $v$ by comparing the subject's assessment $a_{i \in \{\TN,\MI,\PRO\}}(v_j)$ with the assessment provided by the SIG $a_{s\in SIG}(v_j)$ on the same
vulnerability.

% We compare the participant's assessments with the assessments performed by the \CVSS\ SIG for the same vulnerabilities. 

% {\color{red} add part with the formal evaluation}

%\subsubsection{Analysis procedure}\label{sec:mixedreg}

% Students fron the University of Trento performed the assessment between September and November 2014, those from the University of Milan in November 2015, and professionals assessed the vulnerabilities in April 2016.

% The experiments are described following the guidelines by Wohlin et al. \cite{Wohlin-2012-book}. 
% It should be noted that, in this study, a randomize trial cannot be performed, as we cannot randomly assign students to a \MI\ group which requires multiple years of training in security. Our experiment is therefore based on a well-accepted practice, \emph{natural experiment}: different groups with exogenous characteristics are compared and the source of difference and independence is explained. 

%For all groups, the seminar was held by one of the authors and the test was presented as the `hands-on' part of the training. The assessments were completed individually by each participant and were based solely on the vulnerability description with no interactions with instructors.  

We observe subjects in our study multiple times (once per vulnerability). As each
observation is not independent and subjects may learn or understand each vulnerability differently, a formal analysis of our data requires to account for the variance in the observation caused by subject (e.g. rate
of learning or pre-existent knowledge) and vulnerability characteristics (e.g. clarity of description). To
evaluate the effect rigorously, we adopt a set of mixed-effect regression models
that account for two sources of variation: the vulnerability; and the subject~\cite{agresti2011categorical}. The general form of the models is

\begin{equation}
\label{eq:genmod}
 g(y^j_{iv}) = \boldsymbol{x}_{iv}\boldsymbol{\beta} + \boldsymbol{z}_{i}\boldsymbol{u}_{i} + \boldsymbol{h}_{v}\boldsymbol{k}_{v} + \epsilon_{iv},
\end{equation}
where $g(\cdot)$ is the link function, and $y^j_{iv}$ denotes the observation on \CVSS\ metric $j$ performed by subject $i$ on
vulnerability $v$. $\boldsymbol{x}_{ivj}$ is the vector of fixed effects with coefficient $\boldsymbol{\beta}$. The vectors $\boldsymbol{u_{i}}\ \text{and}\ \boldsymbol{k_{v}}$ capture the
shared variability at the subject and vulnerability levels that induces the association
between responses (i.e. assessment error on \CVSS\ metric $j$) within each observation level (i.e. subject $i$ and vulnerability $v$). $\epsilon_{iv}$ is the leftover error. We report regression results alongside a \emph{pseudo-R}$^2$ estimation of the explanatory power of the model for the fixed-effect part as well as for the full model as specified in~\cite{nakagawa2013general}. We report odds ratio (exponentiated regression coefficients) and confidence
intervals (via robust profile-likelihood estimations~\cite{stryhn2003confidence}) for a more immediate model interpretation. Odds lower than one (with $0\leq  C.I.<1$) indicate a significant \emph{decrease} in error rates. These are indicated in Table~\ref{tab:regall},~\ref{tab:regpros} with a $*$ next to the estimate. \emph{Borderline} results are those whose C.I. only marginally crosses the unity up to 5\% (i.e. $0\leq  C.I.\leq1.05$).

\section{Empirical results}\label{sec:analysis}

Our data collection comprises 2190 assessments performed by 73 subjects over 30 vulnerabilities. We consider an assessment as valid if the assessment is a) \emph{complete} (i.e., the whole 
\CVSS\ vector is compiled), and b) \emph{meaningful} (i.e. the assessment is made by assigning a valid value to each \CVSS\ metrics). This leaves us with 1924 assessments for our analysis, or $\approx 88\%$ valid records.
%when appropriate.

%\subsection{Descriptive statistics} \label{sub:sec:descriptive}

\subsection{Effect of security knowledge}\label{sec:secknow}

\subsubsection{Assessment confidence} We start our analysis by evaluating the level of scoring confidence for the three groups for each
vulnerability. 
Table~\ref{tab:confidence} shows the
results for the subjects' reported confidence in the assessments.
\begin{table}[t]
\centering
\small
\caption{Confidence assessments for the groups}
\label{tab:confidence}
\begin{tabular}{l r r r r}
\toprule
&\multicolumn{3}{c}{Confident}&\\
\cmidrule{2-4}
Group & Yes & No & Unsure & tot.\\
\midrule
\TN&228&552&82&862\\
\MI&275&203&57&535\\
\PRO&254&167&106&527\\
tot.&757&922&245&1924\\
\bottomrule
% \MI & 196 & 144 & \textbf{340} \\
% \TN &227 &530 & \textbf{757} \\
% \textbf{tot.} & \textbf{423}&\textbf{674} & \textbf{1097}\\\hline
\end{tabular}
\end{table}
% A Fisher test rejects the null hypothesis of no difference between the groups, and 
% accepts the alternative hypothesis that \MI\ students are more confident in the assessments than \TN\ students
% ($p<0.001$).
Overall, 
 subjects declared to have been confident in their assessment in 39\% (757) 
of the cases, and non-confident in 48\% (922). For the remaining 13\%, 
subjects were unsure. Looking at the different groups, a significant 
majority of scorings in the \TN\ group (64\%) was rated as low confidence, while
 for \MI\ and \PRO\ groups approximately 50\% were confident assessments. 
 Even by considering `Unsure' assessments as low confidence, the figures for the \MI\ and \PRO\ groups are
statistically indistinguishable ($p=1$ for a Fisher exact test\footnote{To avoid issues with dependent observations, we compute the rate of "Yes", "No", "Maybe" answers for each subject, and consider the highest rate to match the subject to a category.}), whereas the difference is significant between \TN\ and \MIPRO\
confidence levels ($p=0.017$).
%This indicates that, whereas there appears to be a clear difference between \TN\ and \MIPRO\ confidence levels ($p=0.017$), this difference disappears once we consider subjects with security knowledge irrespective of their professional expertise.

\subsubsection{Severity estimations}

Whereas technical details may significantly vary between vulnerabilities, for simplicity we grouped the vulnerability assessed into six macro-categories whose definitions have been derived from the \emph{Common Weakness Enumeration} (CWE)\footnote{Details at: \url{http://cwe.mitre.org}, last visited April 2018.}:
%the CWE defines a set of categories to classify vulnerabilities that are similar for type of \emph{weakness} or \emph{flaw} that causes the vulnerability (e.g. a buffer error or flawed authorization mechanism). For sake of presentation, and following OWASP indications\footnote{See \url{https://www.owasp.org/index.php/Category:OWASP_Top_Ten_Project}, last visited April 2018.} we distinguish between vulnerabilities of category:\footnote{A full analysis on the original CWEs returned qualitatively identical results.} 
\begin{itemize}
\item \texttt{input}: vulnerabilities caused by flawed or missing  validation (e.g. code injection);
\item \texttt{information}: vulnerabilities regarding system or process specific (e.g. info disclosure);
\item \texttt{resource access}: vulnerabilities granting the attacker access
to otherwise unauthorized resources (e.g. path traversal);
\item \texttt{crypto}: vulnerabilities affecting cryptographic protocols or systems;
\item \texttt{other}: vulnerabilities that do not belong to specific CWE classes (taken as is from NVD);
\item \texttt{insufficient information}: vulnerabilities for which there is not enough information to provide a classification (taken as is from NVD).
\end{itemize}
%Table~\ref{tab:cwe} details the mapping between CWEs in our dataset and the defined categories.
%\begin{table}[t]
%\centering

% \small
% \caption{Mapping of vulnerability CWE and vulnerability category}
% \label{tab:cwe}
% \begin{tabular}{lp{0.5\columnwidth}}
% \toprule
% Category&CWE\\
% \midrule
% Input&Input validation\\
% Input&Code Injection\\
% Input&SQL Injection\\
% Input&Buffer Errors\\
% Other&Other\\
% Insufficient Information&Insufficient Information\\
% Cryptographic Issues&Cryptographic Issues\\
% Information&Information Leak / Disclosure\\
% Information&Configuration\\
% Resource Access&Improper Link Resolution Before File Access\\
% Resource Access&Permissions, Privileges, and Access Control\\
% Resource Access&Path Traversal\\
% Resource Access&Authentication Issues\\
% \bottomrule
% \end{tabular}
% \end{table}

\begin{figure*}[t]
\centering
\includegraphics[width=0.9\textwidth]{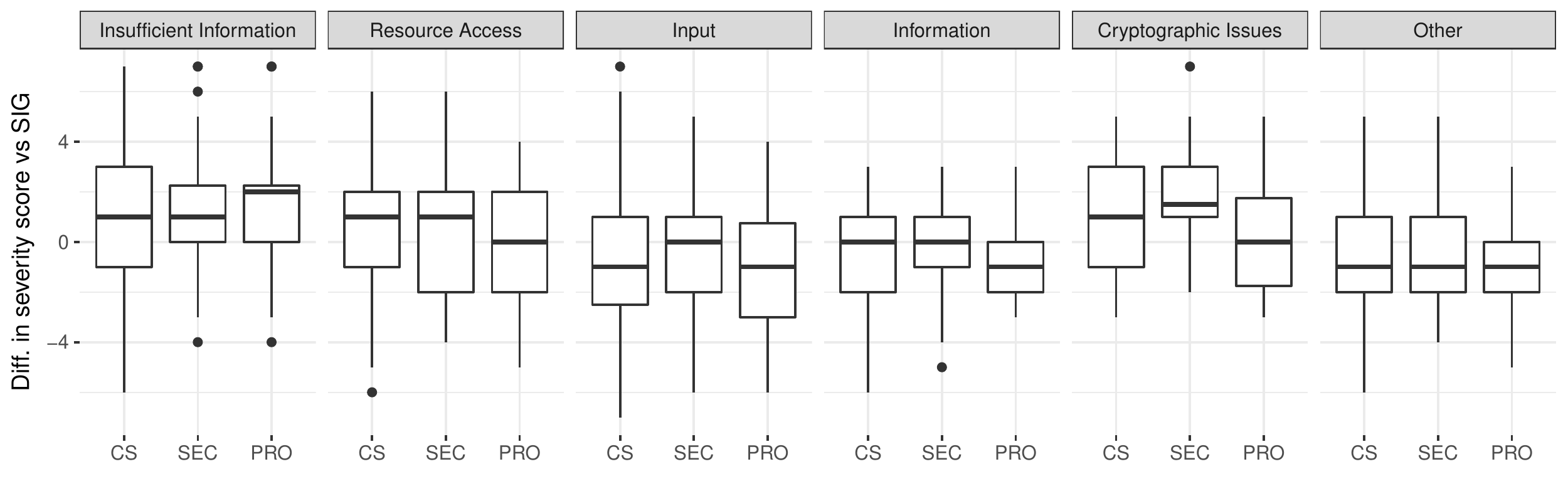}
\begin{minipage}{0.85\textwidth}
\footnotesize
A positive difference indicates that subjects in that group overestimated
the severity of the vulnerability w.r.t. the SIG's score. A negative 
difference indicates underestimation. Groups consistently overestimate vulnerabilities with \texttt{Insufficient information}, and \texttt{Cryptographic issues}.
%, whereas in general subjects of the \PRO\ group appear more aligned with the SIG.
\end{minipage}
\caption{Distribution of difference in severity estimation by vulnerability type and subject group.}
\label{fig:estErrors}
\end{figure*}

Figure~\ref{fig:estErrors}
reports how severity estimations of vulnerabilities vary, w.r.t. the reference score computed by the SIG, between the three groups of participants and for each vulnerability category. A positive difference indicates an \emph{overestimation} (i.e. participants attributed a higher severity score); a negative value indicates an \emph{underestimation}. We observe that \texttt{Cryptographic Issues} and \texttt{Insufficient information} categories were perceived as more severe by all participant groups than by the SIG, whereas for other categories the results are mixed. Following NIST guidelines, and over- or under-estimation of two points may result in an important mis-categorization of the vulnerability, whereas an error of $\pm 0.5$ points is within accepted tolerance levels~\cite{NVD}. Overall, we find that experiment subjects' estimations of vulnerability severity are only marginally off with respect to the SIG estimations.
%In general, however, we find that, on average, disagreements between our groups and the SIG oscillate around \emph{none} with relatively small amplitudes. This indicates that the perceived severity levels are in general well-aligned and do not present significant variations between groups.

\subsubsection{Assessment errors} 
\begin{figure*}[t]
\centering
\includegraphics[width=1.0\textwidth]{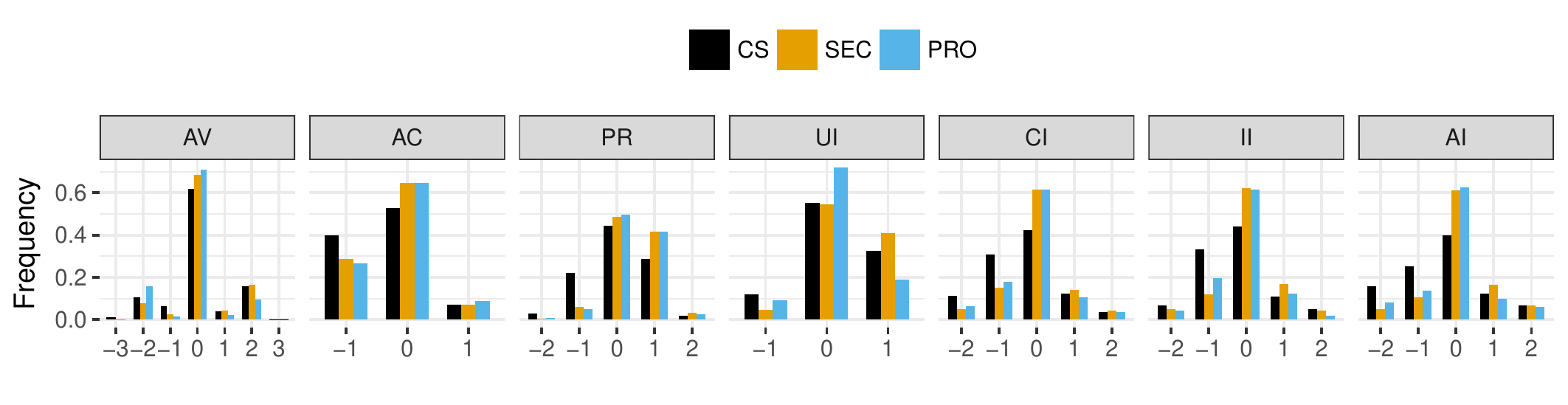}
\begin{minipage}{0.75\textwidth}
\footnotesize
$error = 0$ indicates accordance. $error<0$ indicates that subjects under-estimated
that metric's assessment. $error>0$ indicates that subjects over-estimated it.
\MI\ and \PRO\ subjects are consistently more precise than \TN\ in assessing
vulnerability impact.
\end{minipage}
\caption{Distribution of assessment errors over the 
\CVSS\ metrics.}
\label{fig:errorsImpact}
\end{figure*}

In Figure \ref{fig:errorsImpact}
% \begin{figure*}[t]
% \centering
% \includegraphics[width=0.7\textwidth]{Figures/errors_exploitability.pdf}
% \begin{minipage}{\textwidth}
% \footnotesize
%  $error <0$ indicates that subjects under-estimated
% that metric's assessment. $error>0$ indicates that subjects over-estimated it. Both \MI\ and \TN\ subjects reliably assess \AV. \MI\ seem more precise in assessing \AC\ than 
% \TN. \TN\ seems more precise than \MI\
% at performing \UI\ assessments. The two groups perform comparably over the \PR\
% metric.
% \end{minipage}
% \caption{Distribution of assessment errors for the two groups over the 
% \CVSS\ Exploitability
% metrics. }
% \label{fig:errorsExploitability}
% \end{figure*}
we have a more detailed inspection of scoring errors for the three groups by considering the specific 
\CVSS\ metrics rather than the total score as computed by the \CVSS\ for a vulnerability. We first evaluate the \emph{sign} and the \emph{size} of
errors. 
%As before for the general vulnerability score, errors evaluating single metrics could tend to overestimate or underestimate the correct assessment produced by SIG. 
With regard to the sign of an error, for instance, the \PR metric could have three values (\texttt{High, Low, None}; see Table \ref{tab:CVSSbasemetrics}). Assuming that the SIG attributed the value {\tt Low} for a certain vulnerability, if a participant selects {\tt High} the error is an overestimation (positive error, +1), if he or she selects {\tt None} it is an underestimation (negative error, -1).
%
%by considering an ordered vector of each metric dimension (increasing order) as the difference between the assessment by the subject and the SIG.  This order is provided in Table \ref{tab:CVSSbasemetrics} under column \emph{Values}. 
%A negative error score indicates an \emph{under}-estimation (i.e. the subject
%assigned a \emph{less severe} \CVSS\ value than the correct one); 
%a positive score means an \emph{over}-estimation (i.e. a \emph{more severe} assessment). 
%An error score of \emph{zero} indicates  accordance of the evaluation performed by a subject and SIG. 
%As different \CVSS\ metrics have a different number of possible values, the error size is not on the same scale for each metric.
Errors may also have different sizes, which depend on the specific metric and the specific SIG evaluation. In the previous example, the size of the error is at most 1. However, for a different vulnerability the SIG could have evaluated as {\tt High} for the \PR\ metric. In that case, if a participant selects {\tt Low} it results in a negative error of size 1 (i.e., -1), if he/she selects {\tt None} the error size is 2 (i.e., -2), with different consequences on the overall scoring error for the vulnerability.

Given this computation of errors' sign and size, we observe that the frequency of large errors (defined as errors with size greater then 1), is small. This indicates that, in general, subjects did not `reverse' the evaluation by completely missing the correct answer (e.g. assessing a {\tt High} Confidentiality impact as a {\tt None}), a situation that might have lead to a severely mistaken vulnerability assessment. Whereas a detailed analysis of error margins is outside of the scope of this study we observe that, overall, most subjects in all groups showed a good grasp of the task at hand. 
%Note however that, because of how our experiment is designed, a direct comparison of \emph{absolute performance} between subjects and the SIG is not straightforward to make and no conclusions should be drawn in that regard: the SIG evaluations are based on a wider information set, over a larger timeframe, and approved by voting in the consortium. Differently, in this study the focus is 
%in the comparison between the \TN, \MI, \PRO\ groups, as all subjects followed the same scoring procedure.

The large errors we observe on certain metrics (between 30\% and 60\% of tests, depending on the group of respondents and the metric, as discussed in the following) are mostly produced by errors of size 1. Error rates of this size are to be expected in similar experimental circumstances~\cite[finds error in the 30-40\% rate over a binomial outcome]{onarlioglu2012insights}, particularly considering that participants in our experiment have been explicitly selected with no previous experience in \CVSS\ assessment, the limited amount of time, and the CVE description as the only technical documentation, this rate of small errors is unsurprising. % For these reasons, we believe that results of Figure \ref{fig:errorsImpact} could be used to analyze relative differences in performance between the groups, rather than abilities or effectiveness in absolute terms.
%  
 %Nonetheless, a recent study~\cite{onarlioglu2012insights}, which involved unskilled participants but tested common web application security problems for which participants were familiar with, partially confirms our results, exhibiting errors in the 30-40\% range, which is very much in line with our measurements. On the other hand, a \CVSS\ assessment is composed over a multitude of levels for each metric, rather than boolean true/false answers like in~\cite{onarlioglu2012insights}, which again indicates that the overall performance of our subjects is well in line with what one would expect in general settings.  %Therefore, the error rate of our participants should be weighted with respect to the severity of the error, which can have up to two levels for certain metrics. Considering only severe errors that considerably change the \CVSS\ score with respect to the SIG - i.e., two levels away from the correct answer, for example if the correct answer is {\tt HIGH}, {\tt LOW} is the intermediate level, and {\tt NONE} is the wrong answer selected by the participant - the error rate drops in the range 5-20\%.

 Overall, we observe that there is a clear difference in accuracy between the security unskilled \TN\ and security skilled \MIPRO\ for all metrics. This is particularly evident in the \AV, \AC\ and \PR\ metrics, 
and all \C\I\A\ impact metrics. This effect is also present in the \UI\ metric, but
here the \TN\ and \MI\ students perform similarly, whereas professionals in the \PRO\ 
group achieve higher accuracy. We observe an overall tendency in
\emph{over}-estimating \PR\ and \UI, and \emph{under}-estimating \AC, which
may indicate that relevant information for the assessment of these metrics are missing, a sensible problem already noted in the industrial sector as well (see for example the recent `call for action' from NIST~\cite{nist-VDO}). Conversely, the difference between \MI\ students and \PRO\ professionals seems less pronounced, if present at
all. The tendency of the error does not appear to meaningfully differ between groups, indicating no specific bias toward
over or underestimation.

Table~\ref{tab:estimatePrecisions} reports quantitatively the comparison depicted in Fig.~\ref{fig:errorsImpact}. As error sign does not present obvious between-group differences, for the sake of conciseness here we only consider differences in \emph{absolute} error rates.
\begin{table}[t]
\centering
\small
\caption{Preliminary comparison of group accuracy by \CVSS\ metric.}
\label{tab:estimatePrecisions}
\begin{tabular}{l r r | r r | r r | c c}\toprule
\multicolumn{1}{c}{}&\multicolumn{2}{c}{(1) \TN}&\multicolumn{2}{c}{(2) \MI}&\multicolumn{2}{c}{(3) \PRO}&\multicolumn{2}{c}{Signif.}\\
\cmidrule{2-7}
&err&\multicolumn{1}{c}{sd}&err&\multicolumn{1}{c}{sd}&err&\multicolumn{1}{c}{sd}&1v2&2v3\\
\midrule
% \AV&0.673&0.91&0.563&0.865&0.545&0.871&*&\\
% \AC&0.471&0.499&0.355&0.479&0.353&0.478&**&\\
% \UI&0.447&0.497&0.454&0.498&0.281&0.45&&**\\
% \PR&0.604&0.58&0.551&0.568&0.537&0.564&&\\
% \C&0.722&0.703&0.479&0.661&0.482&0.669&**&\\
% \I&0.675&0.672&0.471&0.658&0.446&0.611&**&\\
% \A&0.824&0.771&0.507&0.697&0.514&0.727&**&\\
\AV&0.67&0.91&0.56&0.86&0.54&0.87&*&\\
\AC&0.47&0.50&0.35&0.48&0.35&0.48&**&\\
\UI&0.45&0.50&0.45&0.50&0.28&0.45&&**\\
\PR&0.60&0.58&0.55&0.57&0.54&0.56&&\\
\C&0.72&0.70&0.48&0.66&0.48&0.67&**&\\
\I&0.68&0.67&0.47&0.66&0.45&0.61&**&\\
\A&0.82&0.77&0.51&0.70&0.51&0.73&**&\\
\bottomrule
% \AV&0.621&0.879&0.684&0.915&\\
% \AC&0.394&0.489&0.473&0.500&**\\
% \UI&0.509&0.501&0.440&0.497&\\
% \PR&0.574&0.567&0.598&0.577&\\
% \C&0.535&0.697&0.708&0.713&***\\
% \I&0.529&0.693&0.663&0.674&***\\
% \A&0.571&0.731&0.839&0.774&***\\\hline
\end{tabular}
\begin{minipage}{0.95\linewidth}\footnotesize
Significance is Holm-Bonferroni corrected and is indicated as: $**$ for $p<0.01$ and $*$ for $p<0.05$ for a Wilcox rank sum test. A more formal evaluation accounting for within-subject dependency of observation is provided in Table~\ref{tab:regall}.
\end{minipage}
\end{table}
At a first approximation, we 
evaluate significant differences between the groups
by employing an unpaired 
Wilcoxon rank sum test,\footnote{This assumes independency between observations. This is violated in our data as the same subject is observed multiple times. Whereas this is formally incorrect, here it only serves the illustrative purpose of quantifying results in Fig.~\ref{fig:errorsImpact}. A formal analysis follows in this Section.} and test the alternative hypothesis
that error size follows the intuition that $e(a_{i\in\PRO}(v_j))<e(a_{i\in\MI}(v_j))<e(a_{i\in\TN}(v_j)),\ \forall j\in \{$\AV, \AC, \UI, \PR, \C, \I, \A$\}$. The statistical significance level is calculated using a Holm-Bonferroni correction
for multiple comparisons. From the results it is apparent that subjects with security knowledge (i.e., \MIPRO)
are significantly more accurate than subjects with no 
security knowledge (i.e., \TN) on all metrics. Interestingly, we find that over all metrics (with
the exception of \UI\ as discussed above), the \PRO\ group does not appear to perform
significantly better than the \MI\ group. Whereas we can not say that the two groups
perform \emph{equally}, it is suggestive to notice that the computed error rates
and standard deviations for the two groups indicate a substantial overlap between the two distributions. 
As discussed in the following, this result should not be considered completely surprising or counterintuitive, rather an interesting convergence with similar experiments performed in different contexts and settings seems to arise, pointing to possible common characteristics of a class of problems.

%This may indicate that the added `professional expertise' that characterizes  the \PRO\ group does not seem to significantly improve the accuracy of the assessment over subjects with security knowledge but limited or no professional expertise. 

%On the other hand, the error rate may be influenced by other factors such as the confidence in the assessment by the subject (e.g. influenced by specific domain knowledge on the vulnerable software), or the type of vulnerability, or unmeasured subject characteristics.
%\footnote{\texttt{Insufficient Information}, and \texttt{Other} match the respective CWE definition that indicates that the vulnerability is either disclosed with insufficient details
%to determine its root cause, or that the vulnerability does not belong to any specific type.}

As each metric has a different set of possible values, to simplify
the interpretation of results,  
we here consider the binary response of \emph{presence} or \emph{absence} of error in the 
assessment. We define
a set of regression equations for each \CVSS\ metric $j$ of the form:

\begin{eqnarray}
\label{eq:reggroup}
g(e_{vi}^j) &=& c + \beta_1 CONF_{vi} + \boldsymbol{\beta_2 GROUP}_{i} \\
\nonumber &+& \boldsymbol{\beta_3 VULNTYPE}_{v} + ..
\end{eqnarray}

where $g(\cdot)$ is the logit link function, $e_{vi}^j$ is the binary response on
presence or absence of error on metric $j$ for subject $i$ and vulnerability $v$,
and $\boldsymbol{\beta_2 GROUP}_{i}$ and $\boldsymbol{\beta_3 VULNTYPE}_{v}$ represent
respectively the vector of subject groups (\TN, \MI, \PRO), and vulnerability
categories.\footnote{We did consider  interaction effects between explanatory variables in the preliminary phases of this analysis, and found qualitatively equivalent results. To avoid complicating the notation and the result
interpretation, we do not report those here.}
% We omit
% error and random variance terms as defined in Eq.~\ref{eq:genmod} for the sake of conciseness.

Table~\ref{tab:regall}
\begin{table*}[t]
\footnotesize
\centering
\caption{Effect of security education on odds of error}
\label{tab:regall}
\begin{minipage}{0.95\textwidth}
\footnotesize
Regression on odds of error accounting for presence or absence of security knowledge
and professional security expertise. \MIPRO\ are significantly more accurate than \TN\
in the assessment. \MI\ does not perform significantly better than \TN\ in the \UI\ metric,
whereas \PRO\ does. Marginal results are obtained for \AV\ and \PR.
\end{minipage}
\begin{tabular}{p{0.2\textwidth} r r r r r r r}
\toprule
error&\multicolumn{1}{c}{\AV}&\multicolumn{1}{c}{\AC}&\multicolumn{1}{c}{\UI}&\multicolumn{1}{c}{\PR}&\multicolumn{1}{c}{\C}&\multicolumn{1}{c}{\I}&\multicolumn{1}{c}{\A}\\
\midrule
c                             & $0.34$           & $1.11$          & $3.26$           & $3.16^{*}$      & $1.01$           & $1.48$          & $0.61$           \\
                              & $[0.11;\ 1.01]$  & $[0.57;\ 2.14]$ & $[0.91;\ 11.75]$ & $[1.06;\ 9.52]$ & $[0.38;\ 2.68]$  & $[0.60;\ 3.66]$ & $[0.22;\ 1.72]$  \\
\MI                           & $0.70$           & $0.58^{*}$      & $1.05$           & $0.75$          & $0.41^{*}$       & $0.46^{*}$      & $0.36^{*}$       \\
                              & $[0.47;\ 1.04]$  & $[0.38;\ 0.87]$ & $[0.72;\ 1.53]$  & $[0.55;\ 1.04]$ & $[0.26;\ 0.64]$  & $[0.32;\ 0.67]$ & $[0.25;\ 0.52]$  \\
\PRO                          & $0.58^{*}$       & $0.59^{*}$      & $0.36^{*}$       & $0.72^{*}$      & $0.39^{*}$       & $0.47^{*}$      & $0.34^{*}$       \\
                              & $[0.39;\ 0.87]$  & $[0.39;\ 0.89]$ & $[0.25;\ 0.53]$  & $[0.52;\ 0.99]$ & $[0.25;\ 0.61]$  & $[0.32;\ 0.68]$ & $[0.23;\ 0.49]$  \\
\texttt{Conf.}                & $0.86$           & $1.00$          & $0.84$           & $1.01$          & $0.71^{*}$       & $0.64^{*}$      & $0.79$           \\
                              & $[0.65;\ 1.11]$  & $[0.78;\ 1.27]$ & $[0.64;\ 1.10]$  & $[0.79;\ 1.28]$ & $[0.55;\ 0.92]$  & $[0.50;\ 0.82]$ & $[0.61;\ 1.01]$  \\
\emph{Vulnerability variables}\\
\texttt{Cryptographic Issues} & $0.43$           & $1.48$          & $0.36$           & $0.17$          & $1.38$           & $1.20$          & $3.74$           \\
                              & $[0.06;\ 2.90]$  & $[0.51;\ 4.32]$ & $[0.04;\ 3.19]$  & $[0.03;\ 1.09]$ & $[0.27;\ 7.08]$  & $[0.27;\ 5.43]$ & $[0.64;\ 21.83]$ \\
\texttt{Information}          & $2.15$           & $1.20$          & $0.19$           & $0.46$          & $2.82$           & $1.53$          & $4.58$           \\
                              & $[0.42;\ 11.21]$ & $[0.47;\ 3.09]$ & $[0.03;\ 1.29]$  & $[0.09;\ 2.47]$ & $[0.66;\ 12.00]$ & $[0.40;\ 5.78]$ & $[0.97;\ 21.87]$ \\
\texttt{Input}                & $2.69$           & $0.67$          & $0.23^{*}$       & $0.50$          & $1.59$           & $0.88$          & $2.91$           \\
                              & $[0.79;\ 9.24]$  & $[0.33;\ 1.35]$ & $[0.05;\ 0.94]$  & $[0.15;\ 1.72]$ & $[0.54;\ 4.66]$  & $[0.32;\ 2.36]$ & $[0.92;\ 9.38]$  \\
\texttt{Resource Access}      & $0.76$           & $0.88$          & $0.18$           & $0.19^{*}$      & $2.22$           & $1.21$          & $4.87^{*}$       \\
                              & $[0.16;\ 3.55]$  & $[0.37;\ 2.11]$ & $[0.03;\ 1.04]$  & $[0.04;\ 0.87]$ & $[0.58;\ 8.54]$  & $[0.35;\ 4.06]$ & $[1.15;\ 20.76]$ \\
\texttt{Other}                & $2.21$           & $0.48$          & $0.08^{*}$       & $0.51$          & $1.68$           & $1.12$          & $4.35$           \\
                              & $[0.42;\ 11.82]$ & $[0.19;\ 1.20]$ & $[0.01;\ 0.53]$  & $[0.10;\ 2.65]$ & $[0.39;\ 7.13]$  & $[0.29;\ 4.25]$ & $[0.92;\ 20.97]$ \\
\midrule
$Var(c|ID)$&0.25&0.33&0.19&0.93&0.38&0.22&0.20\\
$Var(c|CVE)$&1.04&0.30&1.42&0.64&0.79&0.66&0.92\\
$PseudoR^2$ (fixed eff.) & 0.09& 0.04& 0.12& 0.13& 0.07& 0.06& 0.11\\
$PseudoR^2$ (full mod.) &0.34& 0.19& 0.41& 0.41& 0.31& 0.26& 0.34\\
N&1924&1924&1924&1924&1924&1924&1924\\
\bottomrule
\end{tabular}
% \begin{minipage}{0.95\textwidth}
% \footnotesize
% $(*)$ indicates that the C.I. does not overlap $0$.
% \end{minipage}
\end{table*}
reports the regressions' results. We conservatively consider assessments with an `Unsure'
level of confidence (\emph{ref.} Tab.~\ref{tab:confidence}) as non-confident.
Effects for the group variables \MI\ and \PRO\ are with respect
to the baseline category \TN. 
%Negative and significant coefficients indicate 
%a \emph{decreased} chance of error in the presence of the condition.
We report the estimated change in odds of error and confidence intervals of the estimation. 

From the results it appears that {\em subjects with security knowledge, i.e. \MIPRO, 
are significantly more accurate at the assessment than subjects with no 
security knowledge, i.e. \TN, on all metrics}. Overall, \MIPRO\ is between 30\% to 60\%
less likely than \TN\ in making an error. Interestingly, we find that {\em the \PRO\ group does not appear to perform
significantly better than the \MI\ group } (with
the exception of \UI\ metric, as reported in Fig.~\ref{fig:errorsImpact}, for which \PRO\ is approximately 60\% \emph{less likely} to err than \MI\ subjects). A borderline
result is found for the \AV\ and \PR\ metrics, where the C.I. for \MI\ is only marginally crossing 1. This may indicate that the professional expertise that characterizes the \PRO\ group does not necessarily improve the accuracy of the assessment over subjects with security knowledge but limited or no professional expertise.
The effect of confidence on the assessment is relevant for the impact metrics \C\I\A,
indicating that a significant source of uncertainty may emerge from
the effect of the vulnerability on the system. 
Interestingly, we find that some vulnerability types (\texttt{Information} and \texttt{Resource access}) are likely to \emph{induce} error on the \A\ metric, suggesting that specific knowledge or expertise may be needed to discern, for example, between \emph{information} and \emph{service} availability.

Variance by subject ($Var(c|ID)$) and by vulnerability ($Var(c|CVE)$) indicate that the intercept
of the model may vary significantly for each observation (i.e. both different subjects
and vulnerabilities have different `baseline' error rates). This is interesting
in itself as it indicates that neither the subject variables ($\boldsymbol{GROUP}_{i}$)
nor the vulnerability variables ($\boldsymbol{VULNTYPE}_v$), whereas
significant in explaining part of the observed error, may fully characterize the effect.
For example, specific user characteristics or the thoroughness of the vulnerability
description may play a substantial role
in determining assessment accuracy. On this same line, it is interesting to
observe that the overall explicative power of the model is relatively small for all
the considered metrics. This can be expected for random processes 
in natural experiments where the environment can not be fully controlled by the experimenter~\cite{agresti2011categorical} (as exemplified by the variance explained by the full
model as opposed to that of the fixed effects); still, the small $R^2$ values for the fixed effect parameters suggest that the sole presence of security knowledge, even when confounded by assessment confidence and vulnerability type, does not satisfactorily characterize the observation. This further supports that other subject-specific characteristics may drive the
occurrence of an error. We investigate this in the following.

\subsection{Effect of subject characteristics}\label{sec:subjvuln}

To analyze results in finer detail, we use the answers from the questionnaire that 
characterizes \PRO\ subjects as described in Sec.~\ref{sec:methodology}. This allows
us to avoid possible bias in self-reporting by students and to focus
on the target group of the professionals that eventually perform the analysis
in the real world~\cite{Salman-ICSE-15}. 

\begin{figure*}
\centering
\includegraphics[width=0.9\textwidth]{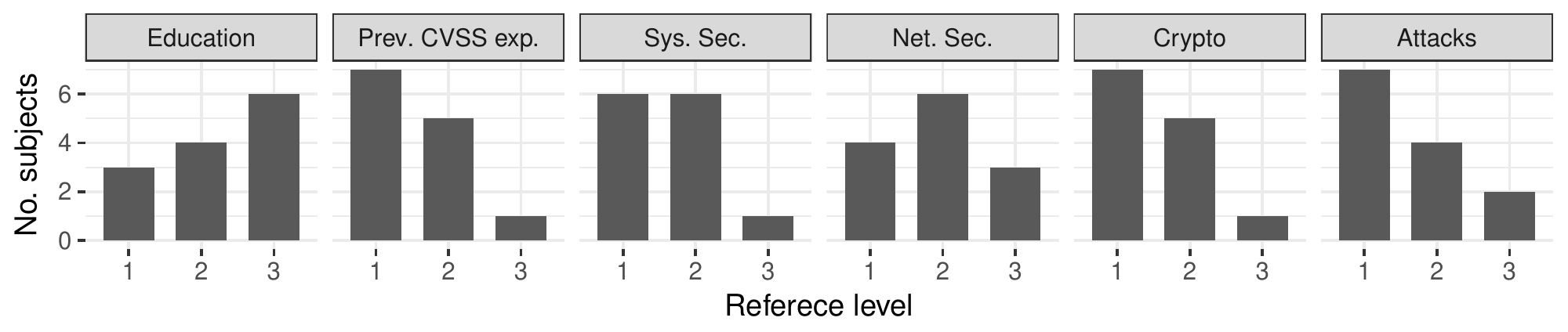}
\begin{minipage}{0.74\textwidth}
\footnotesize
All factors but \emph{CVSS exp.} (nominal) are on an ordinal scale. Reference
levels:
\emph{Education}: 1=High School; 2=BSc degree; 3=MSc degree. \emph{Previous CVSS experience}: 1=None; 2=Yes; 3=NON-CVSS metric. \emph{System security $\rightarrow$ Attacks}: 1=Novice; 2=Knowledgeable; 3=Expert. 
\end{minipage}
\caption{Education and expertise profile of professionals in the \PRO\ group.}
\label{fig:prosdesc}
\end{figure*}

The median subject in the \PRO\ group has six years of professional expertise in the security field, in a range between one and fifteen years ($\mu=5.79,\ \sigma=3.83$).
Figure~\ref{fig:prosdesc}
reports the distribution of the levels for each measured variable. All factors are reported on an ordinal scale (with the exception of \emph{CVSS experience} for
which we have a nominal scale), codified in levels $1\rightarrow3$, where \emph{Education}: 1=High School; 2=BSc degree; 3=MSc degree. \emph{Previous CVSS experience}: 1=None; 2=Yes; 3=NON-CVSS metric. \emph{System security $\rightarrow$ Attacks}: 1=Novice; 2=Knowledgeable; 3=Expert (a fourth level, `None', is not reported as no participant rated him or herself
less than novice on any of these dimensions). Most subjects obtained at least a BSc degree. From discussion during the initial \CVSS\ training it emerged that none of the participants
in the \PRO\ group had a formal specialization in security at the University level.
The group
is evenly split between participants that have previous experience in vulnerability measurement (earlier versions of the \CVSS\ or other methods); most participants rated themselves as `Competent' or `Expert' in Network Security, and are equally split between the levels `Novice' and `Competent or Expert' for all other variables.

To evaluate the effect of subject characteristics on odds of error we first make two considerations: first, the subject distribution seem to be skewed toward presence or absence of expertise or education rather than being meaningfully distributed across
all levels. For example, most subjects attended University with only a handful
interrupting their studies after high school; similarly, few subjects rated themselves
as `experts' in any dimension, with most subjects being either `novices' or `competent'
on the subject matter. We therefore collapse the levels to `novice' or `not novice' to represent this
distinction. Second, some subject characteristics may show high levels of correlation: 
for example, subjects competent in system security may be likely competent on network security as well. Similarly, highly educated professionals may be (negatively) correlated
with years of experience (as more time would be spent on one's studies than on the profession).
We check for multicollinearity problems by calculating the Variance Inflation
Factor of the categorical variables defined above, and 
drop the variables that show evidence of correlation; we keep: \texttt{years, attacks, system security}. We then define the following regression equation:

\begin{eqnarray}
g(e_{vi}^j) &=& c + \beta_1 Years_{i} + \beta_2 Attacks_{i} + \beta_3 SysSec \\
\nonumber &+& \boldsymbol{\beta VULNTYPE}_{v} + ..
\end{eqnarray}

% Table~\ref{tab:cormat}
% \begin{table}
% \small
% \caption{Correlation matrix of \PRO\ characteristics}
% \label{tab:cormat}
% \begin{tabular}{l r r r r r r r}
% \toprule
% &Ed.&Yrs&CVSS.&SysS&NetS&Cryp&Att\\
% \midrule
% 1.&1\\
% 2.&0.41&1\\
% 3.&-0.55&0.01&1\\
% 4.&-0.07&-0.21&0.24&1\\
% 5.&-0.39&-0.27&0.28&0.72&1\\
% 6.&-0.24&-0.32&0.38&0.55&0.62&1\\
% 7.&-0.55&-0.12&0.69&0.24&0.28&0.69&1\\
% \bottomrule
% \end{tabular}
% \end{table}
%reports the results. 

% \begin{figure*}
% \includegraphics[width=0.98\textwidth]{Figures/percerrorsCVE.pdf}
% \caption{Average error rates by CVE and \CVSS\ metric per group}
% \label{fig:CVECVSSGroup}
% \end{figure*}

% To further investigate the effect of professional expertise we evaluate which self-assessed characteristics of participants in the \PRO\ group affect scoring accuracy.
% {\color{red} some descriptive statistic first}
Table~\ref{tab:regpros} reports the results.
\begin{table*}[t]
\centering
\footnotesize
\caption{Effect of subject characteristics on odds of error in the \PRO\ group}
\label{tab:regpros}
\begin{minipage}{0.95\textwidth}
\footnotesize
Regression on odds of error by subject characteristics and vulnerability category. Education, CVSSExp, NetSec, Crypto have been dropped because highly correlated with other factors
in the regression; this is to avoid multicollinearity problems. Overall we find that different vulnerability aspects are covered by different subject characteristics.
\end{minipage}
\begin{tabular}{p{0.2\textwidth} r r r r r r r}
\toprule
error&\multicolumn{1}{c}{\AV}&\multicolumn{1}{c}{\AC}&\multicolumn{1}{c}{\UI}&\multicolumn{1}{c}{\PR}&\multicolumn{1}{c}{\C}&\multicolumn{1}{c}{\I}&\multicolumn{1}{c}{\A}\\
\midrule
c                        & $0.79$           & $2.70$           & $3.42$           & $59.80^{*}$        & $2.22$           & $3.81$           & $0.37$            \\
                         & $[0.14;\ 4.36]$  & $[0.56;\ 13.54]$ & $[0.83;\ 14.79]$ & $[1.83;\ 3027.71]$ & $[0.35;\ 14.13]$ & $[0.43;\ 34.31]$ & $[0.04;\ 2.87]$   \\
\texttt{Years}           & $0.96$           & $0.95$           & $0.86^{*}$       & $0.80^{*}$         & $0.86$           & $0.86$           & $0.90$            \\
                         & $[0.84;\ 1.09]$  & $[0.81;\ 1.11]$  & $[0.76;\ 0.97]$  & $[0.64;\ 0.99]$    & $[0.71;\ 1.05]$  & $[0.68;\ 1.10]$  & $[0.74;\ 1.11]$   \\
\texttt{Attacks}         & $0.49^{*}$       & $0.42^{*}$       & $1.32$           & $0.66$             & $0.45$           & $0.41$           & $0.61$            \\
                         & $[0.26;\ 0.89]$  & $[0.19;\ 0.85]$  & $[0.77;\ 2.25]$  & $[0.24;\ 1.77]$    & $[0.18;\ 1.10]$  & $[0.13;\ 1.23]$  & $[0.23;\ 1.53]$   \\
\texttt{SystemSec}       & $1.14$           & $0.74$           & $0.77$           & $0.74$             & $0.48$           & $0.43$           & $0.42$            \\
                         & $[0.62;\ 2.14]$  & $[0.35;\ 1.54]$  & $[0.44;\ 1.33]$  & $[0.27;\ 2.03]$    & $[0.19;\ 1.20]$  & $[0.13;\ 1.35]$  & $[0.15;\ 1.07]$   \\
\emph{Vulnerability variables}\\
\texttt{Cryp. Issues}    & $0.24$           & $4.67$           & $0.24$           & $0.01$             & $1.20$           & $1.21$           & $6.29$            \\
                         & $[0.01;\ 3.53]$  & $[0.65;\ 42.22]$ & $[0.03;\ 1.79]$  & $[0.00;\ 1.81]$    & $[0.16;\ 9.43]$  & $[0.15;\ 10.07]$ & $[0.59;\ 76.88]$  \\
\texttt{Information}     & $2.13$           & $1.03$           & $0.14^{*}$       & $0.02$             & $1.77$           & $3.05$           & $13.13^{*}$       \\
                         & $[0.23;\ 20.43]$ & $[0.19;\ 5.63]$  & $[0.02;\ 0.83]$  & $[0.00;\ 2.33]$    & $[0.30;\ 10.92]$ & $[0.49;\ 21.10]$ & $[1.64;\ 124.53]$ \\
\texttt{Input}           & $0.81$           & $0.21^{*}$       & $0.17^{*}$       & $0.20$             & $1.08$           & $0.59$           & $4.07$            \\
                         & $[0.15;\ 4.32]$  & $[0.05;\ 0.72]$  & $[0.04;\ 0.62]$  & $[0.00;\ 8.24]$    & $[0.28;\ 4.26]$  & $[0.15;\ 2.40]$  & $[0.82;\ 23.81]$  \\
\texttt{Resource Access} & $0.42$           & $0.62$           & $0.26$           & $0.02$             & $2.25$           & $1.01$           & $15.37^{*}$       \\
                         & $[0.05;\ 3.51]$  & $[0.13;\ 3.00]$  & $[0.05;\ 1.34]$  & $[0.00;\ 1.91]$    & $[0.42;\ 12.57]$ & $[0.18;\ 5.76]$  & $[2.20;\ 131.91]$ \\
\texttt{Other}           & $1.19$           & $0.12^{*}$       & $0.12^{*}$       & $0.23$             & $1.14$           & $0.62$           & $6.13$            \\
                         & $[0.11;\ 12.64]$ & $[0.02;\ 0.74]$  & $[0.02;\ 0.73]$  & $[0.00;\ 37.62]$   & $[0.19;\ 7.10]$  & $[0.09;\ 4.04]$  & $[0.73;\ 57.79]$  \\
\midrule
$Var(c|ID)$&0.02&0.14&0.00&0.51&0.32&0.60&0.34\\
  $Var(c|CVE)$&1.59&0.74&0.83&1.58&0.83&0.90&1.19\\
$PseudoR^2$ (fixed eff.) & 0.07& 0.22& 0.12& 0.14& 0.10& 0.14& 0.16\\
$PseudoR^2$ (full model) &0.38& 0.39& 0.30& 0.47& 0.33& 0.41& 0.43\\
N&357&357&357&357&357&357&357\\
\bottomrule
\end{tabular}
% \begin{minipage}{0.95\textwidth}
% \footnotesize
% $(*)$ indicates that the C.I. does not overlap $0$.
% \end{minipage}
\end{table*}
In general, we observe that not all expertise dimensions are relevant for all evaluation metrics. This is to be expected as, for example, knowledge of in attack techniques may have an impact on evaluating complexity of attack, but may make little difference on other more system-oriented aspects like requirements on user interaction. More in detail, we find that {\em Attack expertise} dramatically decreases error over the \AV\ and \AC\ metrics by almost 60\%. {\em Years of experience} increases
accuracy over the \UI\ and \PR\ metrics (by roughly 20\% per year), explaining the mismatch on \UI\ between
\MI\ and \PRO\ subjects identified in Fig.~\ref{fig:errorsImpact}.
{\em System security} knowledge appears to have a positive impact on the accuracy of assessments on the \C\ and \A\ metrics, but we do not consider this effect to be highly significant.
Results for vulnerability type
are qualitatively equivalent to those reported for the evaluation by group in Tab.~\ref{tab:regall}. Interestingly, the overall explanatory power of the model (accounting for both fixed and random effects) remains satisfactory, and the {\em subject characteristics 
are clearly effective in explaining the variance for most metrics}. The only low ($<10\%) R^2$ fixed-effect values is for \AV\ and can be explained by the low incidence of
error in this metric, which may be then simply be driven by random fluctuations. This is in contrast with the effect for, for example, the \AC\ metric that is characterized by a high variability in error (\emph{ref.} Fig.~\ref{fig:errorsImpact}), and for which 
more than 20\% of the variance is explained by the measured fixed effects. This is in sharp contrast with results in Tab.~\ref{tab:regall} where most of the variance was absorbed by the
random effects.

\section{Discussion}\label{sec:discussion}

%From the experiment's results some observations could be derived.

%This paper discusses how the evaluation of software vulnerability severity according to \CVSS, a formal and structured methodology, changes through individuals with different education and professional profiles. 
%
%The experiment is meant to analyze, first, the difference between those without specific education and experience regarding information security (students of Computer Science) and those with security knowledge (students of Information Security and security professionals), with respect to a problem for which none had a specific training or experience before and requiring a combination of skills and attitudes.

 % from a similar ability to self-evaluating by individuals trained on technical matters to a consequence of the clear structure of \CVSS\ that ease the self-evaluation of the assessor's uncertainty.

%\subsection{Security education and expertise.} 

\textbf{Economics of technical education and of professional experience.}
The value of professional experience and of technical education could be measured by the remunerations offered to IT professionals and through the cost of achieving a degree in security-oriented courses or the cost for an advanced professional training program.     
Therefore, to know that information security knowledge significantly affects the accuracy of a security assessment brings no surprise. 
However, what is almost never measured, and might have important implications in investments for building a competent workforce, is the width of the gain. In other words, the economics of technical education and professional experience is seldom analytically investigated, more often is left to anecdotes or to political discourses.

With our study we have attempted a quantitative estimate of the benefit brought by knowledge. On average, it appears remarkable: those with security knowledge (\MI\ and \PRO\ groups) show error rates reduced by approximately 20\% (see Fig.~\ref{fig:errorsImpact}). 
A second result appears by looking at the average confidence declared by participants: not just assessment accuracy improves with knowledge, but also confidence in assessments. In fact, the unskilled students \TN\ are mostly not confident, while the skilled participants \MIPRO\ declare higher confidence. 
Improving confidence in one's own work is as valuable as improving the accuracy, because it leads to a better self-evaluation, a better control of the task at hand, and optimizes time and efforts.

\textbf{Implications for recruiting and training security professionals.}
What we have seen in our tests is that the combination of skills explains most of the subjects' variance. This is another observation often made anecdotally, but seldom analytically tested in order to be translated into operational policies and tools useful in the definition of recruiting plans or training investments. 
Moreover, competences in different technical domains are correlated. For example, expertise in system security/hardening is highly correlated with expertise in network security; similarly, experience with previous vulnerability assessments and attack expertise go hand in hand.
What emerges from our study is that professionals with same specialization, i.e., cybersecurity, exhibits different competence profiles defined by correlated skills, and different performance in assessing different security metrics are correlated with different profiles. This means that often the specification "security professional with x years of experience", or even a professional qualification like "Security Architect" and the like, could be a too coarse classification to be useful for finding a good match with a security task, especially when the task is mostly oriented to problem solving.

More specific and detailed studies about how specific technical skills correlate and influence problem solving abilities should be useful to inform recruiting and training investments. Recruiting should better know whether specific specialist skills are needed or technical problem solving abilities are required. This should be considered at fine-grained level, by analyzing skills correlation and complementarity. The same holds for professional training, today often unable to be tailored for a clear outcome when a problem solver profile is needed. 

%  suggesting a correlation between vulnerability training and 
% attack/exploitation knowledge. Whereas this comes with perhaps little surprise, it provides an indication of the overlap between security knowledge domains that may serve as a pointer
% for future studies identifying distinctive security `professional profiles'~\cite{reece2015professionalisation} and the possibility to better match professional profiles with tasks' characteristics. 

% A demonstration of matching and mismatching between skills and tasks came from the analysis of the performance of professionals with respect to individual \CVSS\ metrics. 

% We have also observed that most of the variance per user ($Var(c|ID)$ in Tab.~\ref{tab:regpros}) is removed when compared to the same model without the subject characteristics (not reported here for brevity). This indicates that
% the considered skill set of the subjects defines most of the within-subject variability. 
% Further, the considered model covers a significant fraction of the variance when we account 
% for the random effects as well.  

\textbf{Studying socio-technical issues by testing students.}
One clear result from our study is that students with security training and security professionals could perform similarly for some class of problems and under specific conditions. Beside observations related to the economics of technical education and of professional experience, the result is suggesting us another possibility: if we are able to control the conditions leading to similar performance between students and professionals, then for all practical scenarios in which those conditions are realistic, we could reliably study the socio-technical phenomena by testing students, typically easy to enroll in scientific studies, rather than professionals, notoriously very difficult to recruit for experiments.

In general, it seems that similar performance between students and professionals emerges when professionals are unable to address the problem by reusing past mental models or patterns of solution. It also seems to emerge for problems characterized by subjective interpretations and uncertainty about the consequences of an individual evaluation, whereas it disappears in situations where the amount of specialist technical skills or practice gained in years of experience is the only important factor.

We have already encountered such a situation in the software engineering literature. In~\cite{host2000using}, the authors explicitly tested whether CS students could be employed in place of professional programmers for assessing the project's lead time. Similar to our scenario, to perform the assessment, it is required to evaluate subjective data and the importance of contributing factors is uncertain. The study found no significant difference between students and professionals. In~\cite{Salman-ICSE-15}, again, the quality of programming between students and professionals has been tested. The results were that professionals perform significantly better when they could adopt a familiar development approach, but the equivalence between professionals and students returns when an unfamiliar development approach was tested. This effect of experience was also confirmed by cognitive studies~\cite{kalyuga2003expertise}.

Our work add a new scenario regarding information security and problem solving skills. Further experiments could shed more light on the specific conditions that may lead to a substantial equivalence in tests between students with certain knowledge and professionals with given profiles. This would contribute to develop a relevant body of knowledge about the preconditions for defining suitable natural experiments for specific technical aspects, like cybersecurity, without necessarily recruit the scarce and hardly available professionals. 

\textbf{Uncertainty in problem statement affects differently the different groups.}
Another result that consistently emerges from the analysis is that the assessment accuracy may vary significantly among vulnerabilities for all three groups of participants. In the Appendix, specific details and examples are provided.
In general, this represents a confirmation of the bias possibly introduced by framing the security problem. Uncertainty, e.g. produced by an equivocal problem description, might introduce a random error, which may penalize either the experts or the non-experts. 
This suggests that follow-up studies may more formally 
consider the bias introduced by ambiguous vulnerability descriptions on the assessment accuracy and, in general, on problem solving when security threats are affected by uncertainty.

\section{Threats to Validity}

We here identify and discuss the limitations of our study. We consider \emph{Internal and External} threats to validity \cite{Wohlin-2012-book}.

\textbf{Internal}. Subjects in all groups were given an introductory lecture on vulnerability assessment and scoring with \CVSS\ prior to the exercise. 
The exercise was conducted in class by the same lecturer, using the same material.
Another factor that may influence subjects' assessments is the \emph{learning factor}: ``early assessments'' might be less precise than ``late assessments''. All subjects
performed the assessment following a fixed vulnerability order. We address possible
biases in
the methodology by considering the within observation variance on the single vulnerabilities~\cite{agresti2011categorical}. 
Further, the use of students as subjects of an experiment 
can be controversial, especially when the matter of the study is directly 
related to a course that the students are attending\cite{10.1109/METRIC.2003.1232471}. 
Following the guidelines given in 
\cite{Sjoberg-2003-EMSSE,10.1109/METRIC.2003.1232471} 
we made clear to all students  that the exercise is not part of their final evaluation, and that 
 their assessments do not influence their grades or student career.

\textbf{External}. 
% To assess external validity it is important to remark that
% the goal of this study is to evaluate the effect of security expertise on the final assessment of a vulnerability
% and \emph{not} to evaluate \CVSS\ v3 assessments.
% Since we did not use the final \CVSS\ specification language released in June 2015, but rather the one from June 2014's preview document, the actual values of scoring precision might not be generalized to future adoptions
%  of \CVSS\ v3. However, the language used to specify scoring criteria had no influence on our study, as both \MI\ and \TN\ received the same documentation. 
% Although the language of the definitions changed, the metrics' specification released in June 2014 was definitive. 
A software engineer investigating a vulnerability in a real scenario can 
 account for additional information beside the vulnerability description when performing the analysis. This additional information was not provided in the context of our experiment. For this reason we consider our accuracy estimates as conservative (worst-case). Accounting
 for professionals directly addresses this concern and confirms previous
 studies on the representativeness of students in software engineering experiments~\cite{Salman-ICSE-15}. On the other hand, the limited number of participants in the \MI\ and \PRO\ groups, and the difficulty associated with recruiting large
 sets of professionals~\cite{Salman-ICSE-15} calls for further studies on the subject.

 % Finally, it is currently being discussed in the community whether
 % students are representative of professionals in experiments in software
 % engineering \cite{Salman-ICSE-15,10.1109/METRIC.2003.1232471}. However, in this study we are not
 % extending student performances in vulnerability scoring to real-world 
 % professionals; in contrast, we
 % are measuring whether formal security knowledge stemming from University education is a significant factor for vulnerability assessment precision.

\section{Conclusions}

In this study we evaluated the effect of security knowledge and expertise on security vulnerability
assessment. Whereas the case study is specific to the application of \CVSS, it nevertheless provides a useful framework that formalizes adversarial, system, and user perspectives in an
assessment, and therefore well captures the overall skill set on which security practice
is built. 
An important lesson learned with this work is that recurrent effects of expertise and education with respect to some classes of problems described in previous studies unrelated with security, emerge also with an important security task like the vulnerability assessment. Our hypothesis is that the task we have tested in our experiment exhibits some general characteristics common with similar tasks studied, for example, in software engineering contexts. This seems interesting to us since it suggests that possible fruitful approaches to some security problems could be derived from experience in different fields or applying methods and analyses not specifically produced within the security field. 

% We find that whereas formal security knowledge and professional expertise are
% certainly beneficial with respect to absence of any specific security knowledge, this simple distinction does not well capture the effect of the level of specialization of each subject. Hence, the push on a generic `security background' in academia as well as on only certain professionalizing activities (e.g. capture-the-flag, pentesting) may not be adequate in developing a well-trained security workforce. For example, we find that expertise in adversarial techniques helps 
% understanding attack vectors and complexities, but does little to capture user aspects
% or system-specific measures. 
% %These skills seem to be made up for by years of experience in the profession, and may require a deeper focus at the education level. 

Whereas it is not our goal to provide the ultimate guidelines for improving the shortage of well-trained security professionals, some useful lessons might be learned from our study. One is to avoid inflexible assumptions regarding the prevalence of experience over education or adversarial thinking over principles and abstractions. What we have seen is that the ability to face to complex security problems cannot be explained by coarse-grained categories or single characteristics and skill. What matters is always a combination of skills and experience of the assessor and the specifics of the problem to be evaluated. Another lesson learned is that experience may replace formal education, but likewise specific training may provide abilities similar to those brought by experience. Therefore, for both academic education and professional training, the ability to face complex security problems mostly depends on a well-balanced skill set in breadth and depth of knowledge, combined with practical experience.

Future work will replicate this analysis with the final
\CVSS\ language \cite{cvss3}, and include the \emph{Scope}
metric as well. We are particularly interested in cooperating with other researchers
to replicate our study  in different national and educational
contest as results might have important policy implication for 
university education in cybersecurity and eventually for cybersecurity 
in the field.

%\newpage
%\end{document}  % This is where a 'short' article might terminate

%ACKNOWLEDGMENTS are optional
\section{Acknowledgments}
This research has been partly supported from the European Union’s Seventh Framework Programme for research, technological development and demonstration under grant agreement no 285223 (SECONOMICS), and from the NWO through the SpySpot project (no.628.001.004).
%Can be added as submission is not anonymous. 

\bibliographystyle{abbrv}
\bibliography{short-names,security-common,ICSE2016}  % sigproc.bib is the name of the Bibliography in this case

\begin{figure*}
\centering
\includegraphics[width=0.9\textwidth]{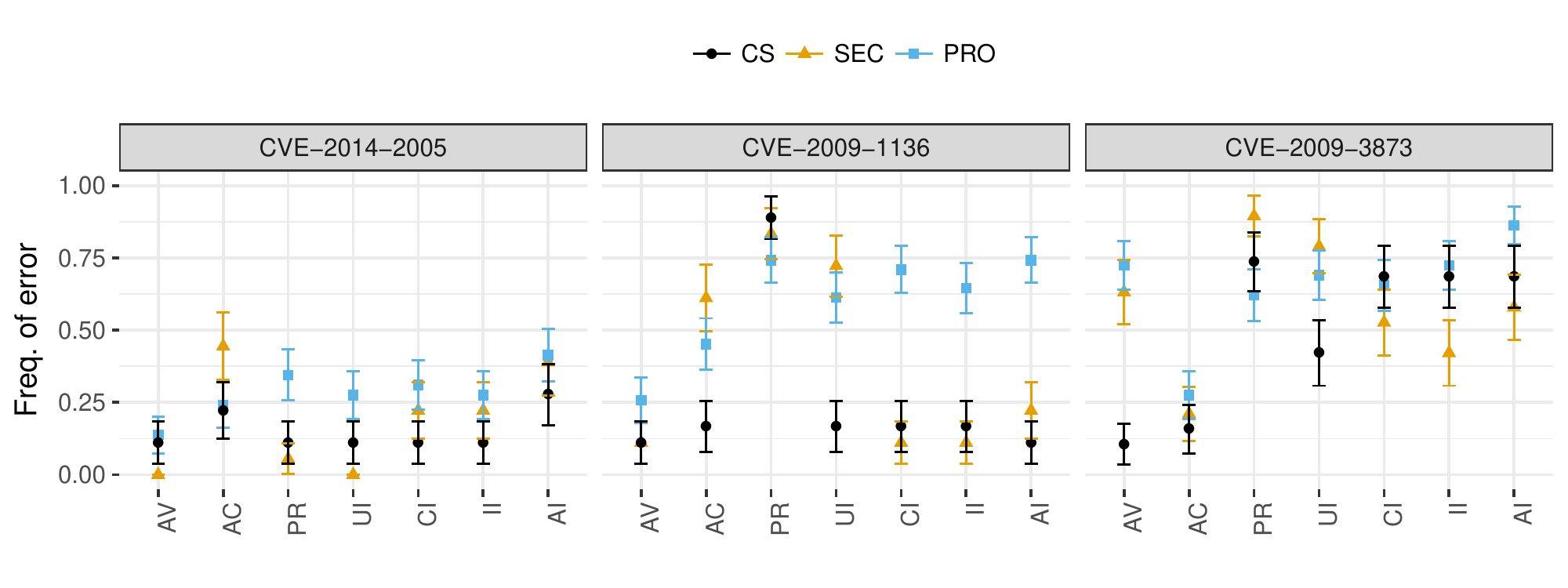}
\begin{minipage}{0.75\textwidth}
\footnotesize
Fraction of erroneous assessments by group for three CVEs. Higher on the scale corresponds to higher error (lower is better). The vertical bars report the standard errors. For the first vulnerability, \texttt{CVE-2014-2005}, the three groups perform similarly over all metrics.
In the second, \texttt{CVE-2009-1136}, security knowledge gives a clear advantage on assessment accuracy, particularly in the \C\I\A\ impact metrics.
Lastly, for \texttt{CVE-2009-3873} we observed mixed results, where security expertise appear to help for \PR, \C\I\A, but not for \AV\ and \AC, with \MI\ performing worse than \TN\ and \PRO\ on the \UI\ metric.
\end{minipage}
\caption{Example of assessment error rates by group on three CVEs}
\label{fig:cvecherry}
\end{figure*}

\begin{figure*}
 \centering
 \includegraphics[width=0.80\textwidth]{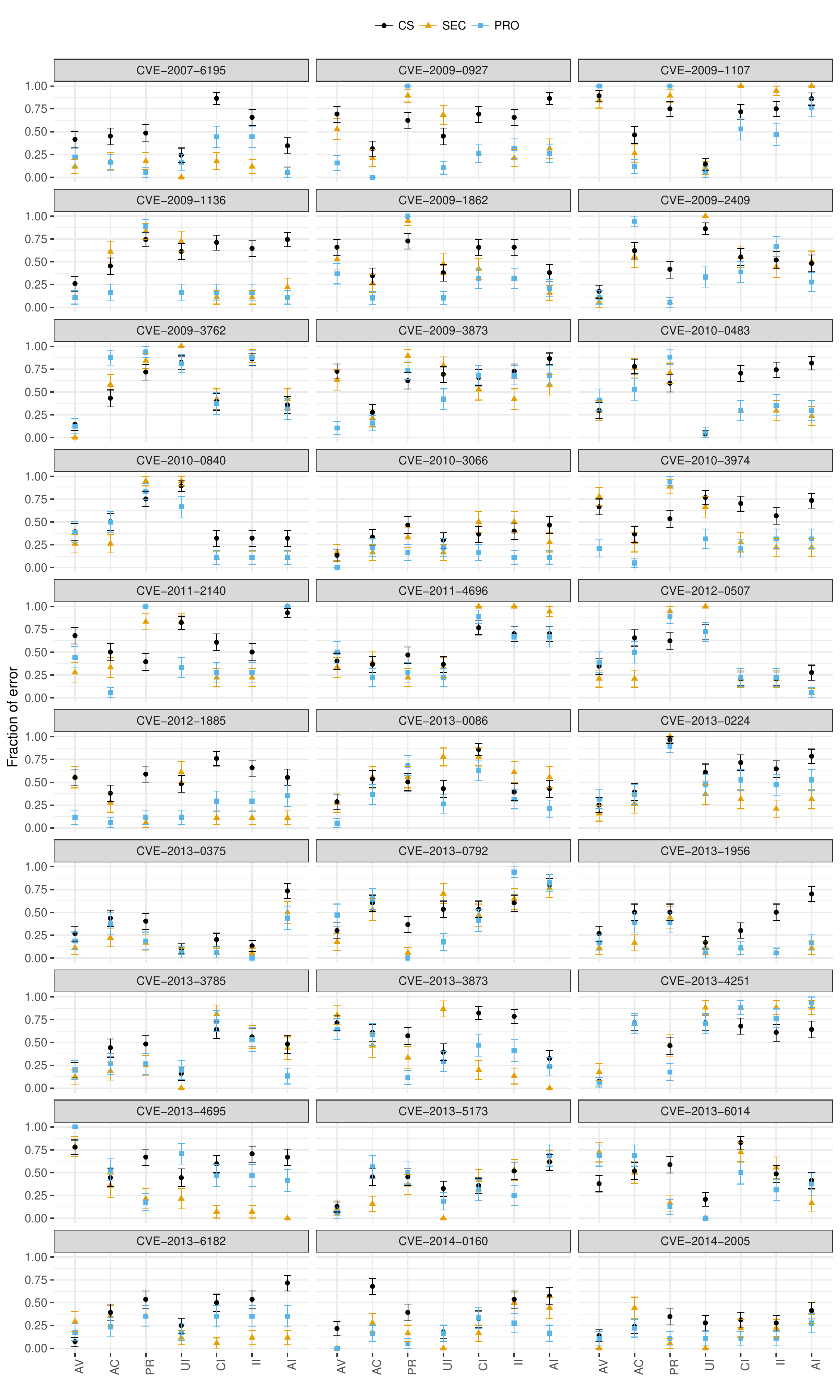}
 \caption{Error rates for \TN, \MI, and \PRO\ by vulnerability and \CVSS\ metrics}
 \label{fig:appcve}
 \end{figure*}

\appendix
Here we discuss the assessment results for three vulnerabilities as examples of the way participants with different skills and experiences have interpreted uncertain information (see Figure~\ref{fig:cvecherry} and the following discussion of the three examples).
Finally, Figure~\ref{fig:appcve} reports error rates for all vulnerabilities of the assessment.
Figure~\ref{fig:cvecherry}
reports the assessment accuracy (expressed in terms
of number of errors) for three  vulnerabilities that represent typical outcomes: (i) the three groups perform similarly; (ii) 
\MIPRO\ have a clear advantage over \TN; (iii) we obtain
mixed results over different metrics. 

\begin{itemize}
\item\emph{Similar accuracy over all metrics (\texttt{CVE-2014-2005})}. %The description for this vulnerability is:
\begin{quote}
\small
Sophos Disk Encryption (SDE) 5.x in Sophos Enterprise Console (SEC) 5.x before 5.2.2 does not enforce intended authentication requirements for a resume action from sleep mode, which allows physically proximate attackers to obtain desktop access by leveraging the absence of a login screen.
\end{quote}
\end{itemize}
%\AV:P; \AC:L; \UI:N; \PR:N; \C:H; \I:H; \A:H\\
From this description, it is clear that the attacker needs to be \emph{physically proximate} to the target system, which gives an obvious clue for \AV; similarly, all groups showed low error rates over the \C\I\A\ assessment,
as it is clear that the attacker gets full (user) access to the system by impersonating the legitimate user. Whereas almost all \MI\ and \PRO\ subjects
understood that the attacker need not be logged in \emph{ahead} of the attack
and scored \PR\ correctly, \TN\ students were likely confused by the existence
of an authentication mechanism for the attacker to bypass. This suggests
that well-formalized security tasks may be accomplished comparably
well by security experts and general IT experts.

\begin{itemize}
\item \emph{Clear effect of security knowledge (\texttt{CVE-2009-1136}).}
\begin{quote}
\small
The Microsoft Office Web Components Spreadsheet ActiveX control (aka OWC10 or OWC11), as distributed in Office XP SP3 and Office 2003 SP3, Office XP Web Components SP3, Office 2003 Web Components SP3, Office 2003 Web Components SP1 for the 2007 Microsoft Office System, Internet Security and Acceleration (ISA) Server 2004 SP3 and 2006 Gold and SP1, and Office Small Business Accounting 2006, when used in Internet Explorer, allows remote attackers to execute arbitrary code via a crafted call to the msDataSourceObject method, as exploited in the wild in July and August 2009, aka "Office Web Components HTML Script Vulnerability."
\end{quote}
\end{itemize}

Whereas all groups correctly understood that the attack can happen remotely (\AV), the security knowledge of \MI\ and \PRO\ has a clear effect on the \C\I\A\ metrics. For this vulnerability, students in both the \MI\ and \TN\
groups were likely confused by the long list of vulnerable systems, giving the impression that these are specific vulnerable software configurations (a criteria for \AC:H~\cite{cvss3}), as opposed to a mere list of vulnerable software. \PRO\ subjects did not get confused by this. In this vulnerability the \PRO\ advantage on the \UI\ metric, discussed in the analysis, is apparent: \PRO\
subjects are the only one that consistently understood that the attack process
requires a user to load a webpage that will \emph{then} load the vulnerable method. This may be easier for \PRO\ subjects to grasp because of the typical
attack dynamics of phishing or XSS attacks commonly received by organizations.
\begin{itemize}
\item \emph{Mixed results (\texttt{CVE-2009-3873}).} 
\begin{quote}
\small
The JPEG Image Writer in Sun Java SE in JDK and JRE 5.0 before Update 22, JDK and JRE 6 before Update 17, and SDK and JRE 1.4.x before 1.4.2\_24 allows remote attackers to gain privileges via a crafted image file, related to a "quantization problem," aka Bug Id 6862968.
\end{quote}
\end{itemize}

The high error for \MI\ and \TN\ students is likely caused by the misleading ``\emph{remote attackers}'' reference in the description: the vulnerability
requires the component to load an image file locally (irrespective of whether this is provided from remote), and qualifies for an \AV:L assessment (see also~\cite[Sec. 3.3 of the User guide]{cvss3}). \PRO\ subjects did not
get tricked by the misleading wording. Again, \PRO\ subjects
outperformed both student groups in the \UI\ metric, understanding that the file need be loaded by the user (e.g. through interaction in a web browser). Interestingly, all groups have a high degree of error in the \C\I\A\ metrics,
suggesting that they deemed ``\emph{gain privileges}'' as a moderate impact,
whereas in most environments Java's JDK/JRE will be running with already high privileges, hence giving the attacker full access.

\end{document}